\documentclass[reprint,nofootinbib,aip,floatfix, rsi]{revtex4-1}
\usepackage[T1]{fontenc}
\usepackage{graphicx}
\usepackage{xspace}
\usepackage{color}
\usepackage{nth}
\usepackage{siunitx}
\usepackage{url}
\usepackage{amsmath}

\DeclareGraphicsExtensions{.pdf,.png,.jpg}

\begin{document}

\preprint{3}

\title{\centering{Large-scale drifts observed on \\electron temperature
  measurements on JET plasmas}}

\author{Thomas \surname{Gerbaud}}
\email[]{Permanent e-mail address: tgerbaud@gmail.com}
\affiliation{JET-EFDA, Culham Science Centre, Abingdon, OX14 3DB, UK}

\author{Stefan \surname{Schmuck}}
\affiliation{Max-Planck-Institut für Plasmaphysik, Teilinsitut Greifswald, EURATOM-Assoziation, D-17491 Greifswald, Germany}

\author{Barry \surname{Alper}}
\author{Kieran \surname{Beausang}}
\author{Marc \surname{Beurskens}}
\author{Joanne \surname{Flanagan}}
\author{Mark \surname{Kempenaars}}
\author{Antoine \surname{Sirinelli}}
\affiliation{Euratom/CCFE Fusion Association, Culham Science Centre, Abingdon, Oxon, OX14 3DB, UK}

\author{Mikhail \surname{Maslov}}
\affiliation{Association EURATOM-Confédération Suisse, Ecole Polytechnique Fédérale de Lausanne (EPFL), CRPP, CH-1015 Lausanne, Switzerland}

\author{Guilhem \surname{Dif-Pradalier}}
\affiliation{Association EURATOM-CEA, CEA/DSM/IRFM, Cadarache 13108 Saint Paul Lez Durance, France}

\author{JET EFDA Contributors}
\altaffiliation{See the Appendix of F. Romanelli et al., Proceedings of the 23rd IAEA Fusion Energy Conference 2010, Daejeon, Korea}
\affiliation{JET-EFDA, Culham Science Centre, Abingdon, OX14 3DB, UK}

\date{\today}


\newcommand{\nameMI}   {MICH}
\newcommand{\nameLIDAR}{LIDAR}
\newcommand{\nameECE}  {HRAD}
\newcommand{\nameHRTS} {HRTS}

\newcommand{\mi}   {{\textsc{\footnotesize \nameMI}}\xspace}
\newcommand{\lidar}{{\textsc{\footnotesize \nameLIDAR}}\xspace}
\newcommand{\ece}  {{\textsc{\footnotesize \nameECE}}\xspace}
\newcommand{\hrts} {{\textsc{\footnotesize \nameHRTS}}\xspace}
\newcommand{\efit} {{\textsc{\footnotesize EFIT}}\xspace}

\newcommand{\ecem}{{\mathrm{\scriptscriptstyle \nameECE}}}
\newcommand{\hrtsm}{{\mathrm{\scriptscriptstyle \nameHRTS}}}

\newcommand{\Tem}{\mathrm{T_{e}}}
\newcommand{\Te}{$\Tem$\xspace}
\newcommand{\Nem}{\mathrm{n_{e}}}
\newcommand{\Ne}{$\Nem$\xspace}

\newcommand{\Temim}{\mathrm{\Tem_{\scriptscriptstyle \nameMI}}}
\newcommand{\Telidarm}{\mathrm{\Tem_{\scriptscriptstyle \nameLIDAR}}}
\newcommand{\Teecem}{\mathrm{\Tem_{\scriptscriptstyle \nameECE}}}
\newcommand{\Tehrtsm}{\mathrm{\Tem_{\scriptscriptstyle \nameHRTS}}}
\newcommand{\Temi}{$\Temim$\xspace}
\newcommand{\Telidar}{$\Telidarm$\xspace}
\newcommand{\Teece}{$\Teecem$\xspace}
\newcommand{\Tehrts}{$\Tehrtsm$\xspace}

\newcommand{\rhoecem}{{\rho_{\ecem}}}
\newcommand{\rhohrtsm}{{\rho_{\hrtsm}}}
\newcommand{\recem}{{r_{\ecem}}}
\newcommand{\rhrtsm}{{r_{\hrtsm}}}

\newcommand{\field}{$\mathrm{B_0}$\xspace}
\newcommand{\Ip}{$\mathrm{I_P}$\xspace}
\newcommand{\Pohm}{$\mathrm{P_{OHM}}$\xspace}
\newcommand{\Ptotm}{\mathrm{P_{ADD}}}
\newcommand{\PICRH}{$\mathrm{P_{ICRH}}$\xspace}
\newcommand{\PICRHm}{\mathrm{P_{ICRH}}}
\newcommand{\Ptot}{$\Ptotm$\xspace}
\newcommand{\PNBIm}{\mathrm{P_{NBI}}}
\newcommand{\PNBI}{$\PNBIm$\xspace}
\newcommand{\li}{$\mathrm{l_i}$\xspace}
\newcommand{\Wdiam}{\mathrm{W_{DIA}}}
\newcommand{\Wdia}{$\Wdiam$\xspace}
\newcommand{\Wthm}{\mathrm{W_{th}}}
\newcommand{\Wthmi}{\mathrm{W_{th,i}}}
\newcommand{\Wthme}{\mathrm{W_{th,e}}}
\newcommand{\Wth}{$\Wthm$\xspace}
\newcommand{\Wpetm}{\mathrm{W_{fast,\perp}}}

\newcommand{\Zeffm}{\mathrm{Z_{EFF}}}
\newcommand{\Zeff}{$\Zeffm$\xspace}
\newcommand{\rangePtot}[2]{${#1}\!<\!\Ptotm\!<\!{#2}$ MW\xspace}
\newcommand{\mus}[1]{$#1{\mu}s$\xspace}

\newcommand{\maxTem}{\mathrm{\Tem_{\scriptscriptstyle MAX}}}
\newcommand{\maxTe}{$\maxTem$\xspace}
\newcommand{\intTe}{$\intTem$\xspace}
\newcommand{\intTem}{\mathrm{\Tem_{\scriptstyle {\scriptscriptstyle-}0.4 \mapsto 0.9}}}
\newcommand{\TeFix}[1]{$\mathrm{\Tem_{#1}}$\xspace}

\newcommand{\maxTeECEm}{\mathrm{\Tem_{\scriptscriptstyle MAX,\mi}}}
\newcommand{\maxTeLIDARm}{\mathrm{\Tem_{\scriptscriptstyle MAX,\lidar}}}

\newcommand{\maxTeECE}{$\maxTeECEm$\xspace}
\newcommand{\maxTeLIDAR}{$\maxTeLIDARm$\xspace}

\newcommand{\TeECE}{\mathrm{\Tem_{\scriptscriptstyle \mi}}}
\newcommand{\TeLIDAR}{\mathrm{\Tem_{\scriptscriptstyle \lidar}}}

\newcommand{\kcd}{\texttt{KC1D}\xspace}

\newcommand{\median}{$\text{median}$}
\newcommand{\JPN}{\text{JPN}\xspace}

\newcommand{\Part}[1]{\ref{#1}}
\newcommand{\Fig}[1]{Fig.~\ref{Fig#1}}
\newcommand{\Tab}[1]{Table~\ref{Tab#1}}

\newcommand{\mytilde}{\raise.17ex\hbox{$\scriptstyle\sim$}}
\newcommand{\ie}{i.e.\xspace}
\newcommand{\eg}{e.g.\xspace}
\newcommand{\mypm}[1]{\hbox{$\pm$#1\%}}
\newcommand{\mypmr}[2]{\hbox{$\pm$#1-#2\%}}

\begin{abstract}

  Between 1995 and 2009, electron temperature (\Te) measurements of more
  than 15000 plasmas produced in the Joint European Torus (JET) have been
  carefully reviewed using the two main diagnostics available over this
  time period: Michelson interferometer and Thomson scattering systems. Long
  term stability of JET \Te is experimentaly observed by defining the
  ECE TS ratio as the ratio of central \Te measured by \mi and \lidar.

  This paper, based on a careful review of \Te measurement from 15 years
  of JET plasmas, concludes that JET \Te exhibits a 15-20\% effective
  uncertainty mostly made of large-scale temporal drifts, and an overall
  uncertainty of 16-22\%.

  Variations of 18 plasma parameters are checked in another data set,
  made of a \textit{reference data set} made of ohmic pulses as similar
  as possible between 1998 and 2009. Time drifts of ECE TS ratios appear
  to be mostly disconnected from the variations observed on these 18
  plasma parameters, except for the very low amplitude variations of
  \field which are well correlated with off-plasma variations of a
  8-channel integrator module used for measuring many magnetic signals
  from JET.

  From mid-2002 to 2009, temporal drifts of ECE TS ratios are regarded
  as calibration drifts possibly caused by unexpected sensitivity to
  unknown parameters; the external temperature on JET site might be the
  best parameter suspected so far.

  Off-plasma monitoring of \mi made of calibration performed in the
  laboratory are reported and do not appear to be clearly correlated
  with drifts of ECE TS ratio and variations of \kcd
  integrators. Comparison of estimations of plasma thermal energy for
  purely Ohmic and NBI-only plasmas does not provide any definite
  information on the accuracy of \mi or \lidar measurements.

  Solutions aiming at tracking down these unexpected uncertainties of
  JET \Te are detailed and can be performed during next JET campaigns
  (C28+, after October 2011), for instance with highly-reproducible
  reference pulses and off-plasma monitoring of the diagnostics.

  Whatever causes these \Te drifts, this experimental issue is regarded
  as crucial for JET data quality.

\end{abstract}

\pacs{}

\maketitle 

\section{Introduction}

In this paper 15 years of electron temperature (\Te) measurements made
by multiple diagnostics on plasmas performed in the Joint European
Torus (JET) have been analysed and carefully reviewed.

JET offers the unique possibility to assess \Te stability and, then, its
effective uncertainty, by checking the very long-term stability of
electron temperature measurements provided by multiple diagnostics on a
large sized tokamak. Lessons learn from this data review are then
expected to provide solid feedback for JET operation and benefit future
ITER operation.

In \Part{Sec:DataDiagnostics_diags}, diagnostics measuring the electron
temperature profile on JET are described: Michelson interferometer
(\mi), core Thomson scattering (\lidar), ECE heterodyne radiometer
(\ece) and the high resolution Thomson scattering
system (\hrts). Some details about the absolute calibration of the ECE
diagnostics (\mi, \ece) are given in (\Part{Sec:DataDiagnostics_ECE_cal}).
More than fifteen years of JET data are analysed in this paper, with
different time and space resolution and non-negligible data scattering
(\Part{Sec:DataDiagnostics_data}). The way  \Te data from different
diagnostics are compared is stated in (\Part{Sec:DataDiagnostics_mapping})
for the magnetic mapping of the diagnostics' line of sight. JET \Te
observables are defined in (\Part{Sec:DataDiagnostics_meas}).


In \Part{Sec:Obs1}, experimental observations of the long term stability
of JET \Te between 1995 and 2009 are done by using the ECE TS ratio
(\Part{Sec:Obs1_ECE_TS_ratio}). Two data sets are used in this paper
(\Part{Sec:Obs1_sets}): the \textit{large data set}, from 1995 to 2009,
where \mi and \lidar data are available and with a broad selection of
plasma parameters (\Part{Sec:Obs1_wide_95_09}); and the \textit{reference
  data set} made of ohmic pulses as similar as possible between 1998 and
2009, including \Te data from the four diagnostics
(\Part{Sec:Obs1_ref_0809}). Reduced set limited to the latest JET
campaigns is used as well, ranging from campaign C20 (April 2008) to C27
(November 2009) \ie JET pulse numbers (\JPN) ranging from 72639 to 79853.

In \Part{Sec:Obs2}, the evolution patterns and/or instrumental drifts of
ECE TS ratio are linked with small variations of the magnetic field, not
caused by or linked to JET plasmas, namely off-plasma. 

In \Part{Sec:Obs3}, the focus is set on \mi as off-plasma monitoring
has been carried out in the laboratory since 1984 -- this is used to assess the
stability of the diagnostic. Data from 1998 to 2010 are presented and
compared with the time drifts of the ECE TS ratio.

In \Part{Sec:Wth}, the plasma thermal energy (\Wth) is estimated from a
kinetic expression, that is using \Te profiles measured by \mi and
\lidar, then compared to \Wth from diamagnetic measurements.

In \Part{Sec:Obs4}, the profile mismatches of high resolution
measurements from ECE (\ece) and TS (\hrts) systems are
documented. Links are made with ECE TS ratio.

The findings of the study are then discussed in (\Part{Sec:Dis}), where
the overall uncertainty of JET \Te measurements between 1995 and 2009 is
assessed. Comments and recommendations are then provided and solutions
to track down the unexpected variations of the ECE TS ratio are
detailed.

\section{Diagnostics \& Data}
\label{Sec:DataDiagnostics}

\subsection{Diagnostics}
\label{Sec:DataDiagnostics_diags}

In the JET tokamak, radial profiles of electron temperature are measured
by diagnostics based on electon cyclotron emission (ECE, Refs
\cite{costley-PRL-74, wesson, hutchinson_book}): Michelson
interferometer (\mi), heterodyne radiometer (\ece); and based on Thomson
scattering (TS, Refs \cite{wesson, hutchinson_book}): \lidar and a
high-resolution TS system (\hrts). \mi and \lidar are both absolutely
calibrated while \ece is cross-calibrated. Even if \hrts could be
independently calibrated, \hrts data used in this article are actually cross-calibrated. Main
characteristics of these diagnostics are summarized in \Tab{0}.

The Michelson interferometer is originally described in
Ref~\cite{costley-EC4} in its original 1983 set-up; a description of the
current (2012) set-up can be found in Ref~\cite{schmuck-2012}. The
absolute calibration of \mi used in this study has been performed in
1996 using the usual cold/hot sources technique \cite{baker-EC4-cal,
  Bartlett-EC6} based on Rayleigh-Jeans' law. Such a calibration
requires the positioning of sources in front of the diagnostic in-vessel
antenna: the interferometer can then be calibrated against these
sources, characterised by their surface temperature, emissivity and
frequency domain. \mi is calibrated for the
\SIrange{50}{350}{\giga\hertz} frequency domain, with a 10\%
uncertainty.

The \ece diagnostic is a heterodyne radiometer installed in 1987
\cite{salmon-EC6} in a 8-channel configuration covering the
\SIrange[range-phrase=-,range-units=single]{73}{79}{\giga\hertz}
range. Constantly upgraded (see Refs~\cite{bartlett-EC8,
  delaluna-RSI-04, barrera_ppcf_10} and references herein), 96 channels
are available since 2008, covering
\SIrange[range-phrase=-,range-units=single]{69}{207}{\giga\hertz}
range. \ece is cross-calibrated against \mi for each JET pulse by equalising
signals from each radiometer channel to the temperature measured by \mi
for the same ECE frequency \cite{zerbini-EC-12}. A fully calibrated
radiometer is technically not out of reach but absolute calibrations
should be performed after each modification or upgrade. Furthermore
high-frequency microwave components ($>$\SI{100}{\giga\hertz}) are quite
prone to failure, strongly reducing the effectiveness of such a
calibrated radiometer.

The \lidar \cite{gowers-RSI-95} is a Thomson Scattering diagnostic based
on a time-of-flight technique, first installed in 1987, where radial
position is determined by the time it takes the scattered light to reach
the spectrometer. In the spectrometer the incoming light signal is
divided into several spectral intervals
(\SIrange{370}{850}{\nano\meter}) by interference filters and then
focused into fast photosensitive detectors. Calibration of the system is
performed in several steps. First the spectral response of each
individual channel is measured by illuminating the spectrometer input
slit with a monochromatic light source with adjustable wavelength. The
whole collection system is illuminated by a calibrated white light
source placed in front of the JET vacuum windows and the relative
sensitivity of each channel is recorded. Spectral transmission of the
windows is measured separately and applied as a correction to the
calibration factors. Window transmission is chromatic (depends on
wavelength) and the associated correction does cause a change in
measured \Te but not more than 5\%. Window transmission was measured
several times during 2005-2011 and appeared to not change
significantly. The spectral calibration is relatively easy to perform
and tends to show the same results every time, and as has been stable
for many years ($>15$). Systematic error in \Te
measurements caused by these calibration uncertainties is very unlikely to
exceed 5\%. Position calibration is performed by observing the light
scattered from the laser beam dump with well known location at the JET
inner wall.

The \hrts diagnostic\cite{Pasqualotto-RSI-2004} measures the electron
temperature (\Te) and density (\Ne) along a chord that runs near the
plasma mid-plane, from the plasma core to the outer pedestal
region. This profile is measured by analysing the light that is Thomson
scattered from a \SI{3}{\joule}, \SI{20}{\hertz}, Q-switched Nd:YAG
system operating at \SI{1064}{\nano\meter}. The scattered light is
coupled into a linear array of more than 130 optical fibers which feed
into a bank of 21 polychromators. Each polychromator analyzes three
separate signals, via delay-line multiplexing, allowing a maximum of 63
spatial points in the \hrts profile. The spectral response and absolute
transmission level for each spatial point is independently
calibrated. The absolute position of the profile is performed by
back-illuminating the optical fiber array onto an in-vessel ruler. Due
to (corrected for data acquired after 2010) problems with the spectral
calibration process, the \hrts \Te data from JET campaigns  for C23-C27 (\JPN=74391-79853) relies
in part on cross-calibration with the \ece diagnostic
\cite{flanagan-cumbria-10}. This cross-calibration is performed by
comparing the \Te profile with the \ece diagnostic for a handful of
carefully selected dedicated calibration pulses. The resultant
calibration correction factors are then validated by comparing the
corrected profiles over many 1000's of JET pulses.

\begin{table}[htbp]
  \begin{center}
    \begin{tabular}{ c|c|c|c}
      \hline
      Diagnostic & measurement  & measurement & radial     \\
                 & time         & frequency   & definition \\
      \hline
      \mi & 
      \SI{17}{\milli\second} & 
      \SI{30}{\hertz} &
      \mytilde \SI{10}{\centi\meter} 
      \\
      \ece & 
      \SI{200}{\micro\second}  & 
      \SI{5}{\kilo\hertz} &
      \SIrange[range-phrase=-,range-units=single]{2}{5}{\centi\meter} 
      \\
      \lidar & 
      \SI{5}{\nano\second} & 
      \SI{4}{\hertz} &
      \mytilde \SI{12}{\centi\meter} 
      \\
      \hrts & 
      \SI{20}{\nano\second} & 
      \SI{20}{\hertz} &
      \mytilde \SI{1}{\centi\meter} 
      \\
      \end{tabular}
      \caption{ \label{Tab0} Temporal \& radial characteristics of JET
        diagnostics measuring \Te}
      \end{center}
\end{table}

\begin{figure}
\includegraphics[width=.8\linewidth]{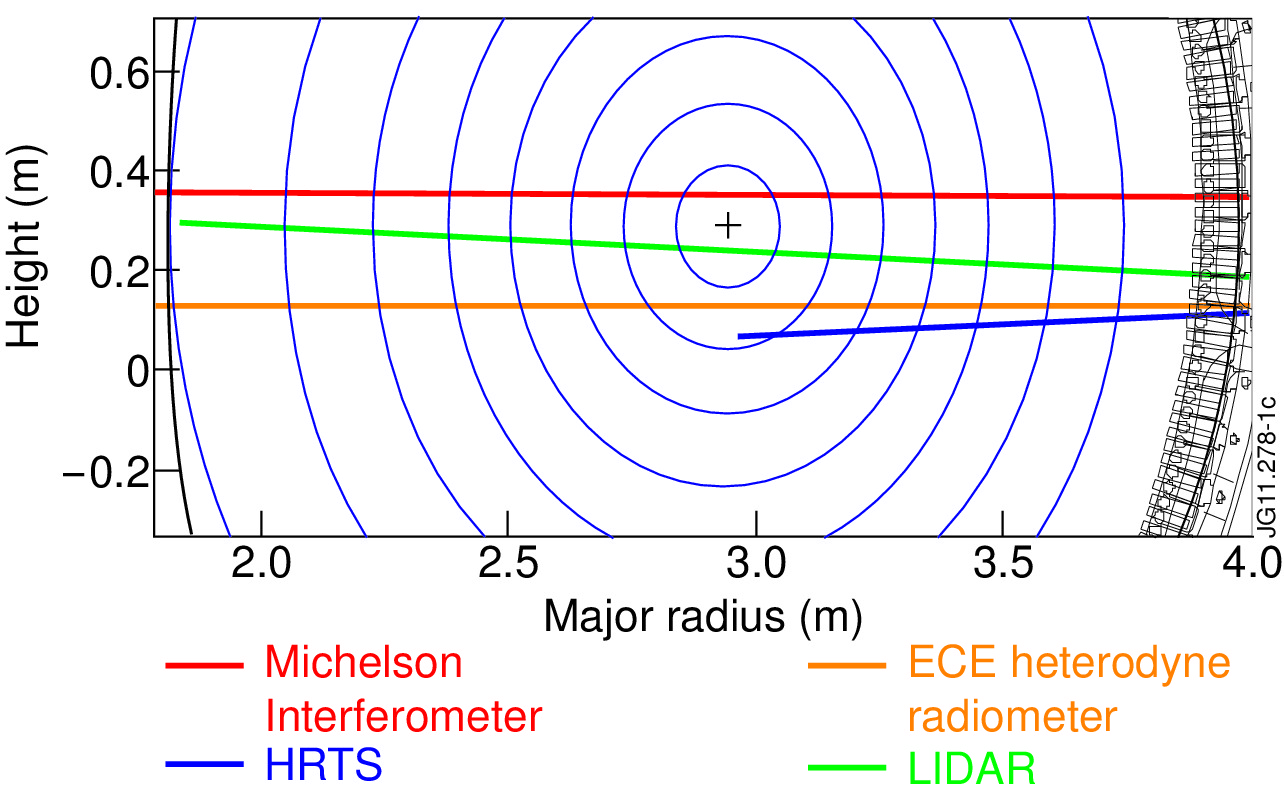}
\caption{\label{Fig1} Line-of-sights of \mi, \ece, \lidar and \hrts for
  typical ohmic plasma (\JPN 79853, t=5.0s). Flux surfaces are computed by
  \efit.}
\end{figure}

The diagnostics' lines-of-sight (LOS) are plotted in \Fig{1}. For vessel
configurations used between 1995 and 2009, \mi and \lidar LOS are
usually very close to the plasma midplane by a few
\si{\centi\meter}. \ece and \hrts LOS are situated at lower z-position
and usually miss the plasma centre. Mapped on the plasma midplane, \ece
\Te profiles usually exhibit a gap for $|\rho|<0.2$ and \hrts ones are
restricted to the low field side ($\rho>0.2-0.3$), where $\rho=\pm\sqrt{\phi}$ and
$\phi$ the normalised poloidal flux.


\subsection{About ECE calibrations}
\label{Sec:DataDiagnostics_ECE_cal}

\mi's absolute calibration is used by \ece and \hrts (for C20-C27); these
diagnostics provide the JET \Te radial profiles with high time and
radial resolutions that are usually shown in publications. 

Absolute calibration of \mi follows, for each frequency $F$:
\begin{equation*}
  T_{\mathrm{keV}}(F)=I_{\mathrm{meas}}(F)/a(F)
\end{equation*}

with $I_{\mathrm{meas}}(F)$ the uncalibrated intensity measured by \mi and
$a(F)$ the F-dependent calibration factors.

A radial calibration of the ECE system requires a magnetic reconstruction code
(\efit\cite{Lao_NF_85, obrien_NF_92}) that links the frequency domain of
the observed ECE to the corresponding radial domain. The usual
assumption of a cold ECE resonance is made, which means that $\omega_n =
n q_e B_0/m_e$ \ie all broadening effects, including relativistic ones,
are neglected; see Ref~\cite{bekefi_book} for a detailed analysis. This
assumption results in a slight (a few \si{\centi\meter}) artificial
shift of the ECE \Te profiles in the outwards direction
($\mathrm{R_{ECE,\neq cold} < R_{ECE,cold}}$).

In order to update the (undocumented) 1996 ECE calibration in use for
data measured between 1995 and 2010, a calibration campaign has been
performed in 2007\cite{KK1_zerbini_2007} and repeated for confirmation in
2010\cite{KK1_schmuck_2011}, addressing most of the issues raised in the
conclusion of the 2007 campaign.

Both calibration campaigns indicate a small decrease of \mi's
sensitivity in the relevant ECE range (\SIrange{90}{220}{\giga\hertz}),
causing (if applied) a $15-20\%$ increase of JET \Te. Through
cross-calibrations procedures (see (\Part{Sec:DataDiagnostics_diags})),
\ece and \hrts measurement would be affected similarly. 

Following this revaluation, the periods of validity of the 1996, 2007
and 2010 ECE calibrations have been assessed. Data from \mi, \ece and
\hrts presented in this study still use the 1996 calibration. Applying
the 2010 ECE calibration leads to a noticeable increase of \Te affecting
the ECE TS ratio but not its variations.


\subsection{Data}
\label{Sec:DataDiagnostics_data}

Data analysis mostly focuses on \mi and \lidar measurements, which might
be highly scattered: $\mypm{5}$ for \mi and $\mypmr{10}{15}$ for \lidar.
Time averaging is then required. JET pulses have been cut into
timeslices of \SI{1}{\second} length where the total field (\field),
plasma current (\Ip), additional power (\Ptot), electron temperature and
density (\Te, \Ne) are kept reasonably constant: $\leq 1\%$ variations
for \field, $\leq 5\%$ for \Ip, $\leq 10\%$ for central density,
diamagnetic energy and additional power.  All the plasma measurements
presented in this paper are extracted from the JET public database.

In order to improve the readabiliy of the figures, temporal axes
often represent timeslices instead of pulse dates or \JPN. Timeslices
selection is roughly linear with \JPN as pulses usually have 5 to
10 timeslices with ohmic conditions, and not linear with pulse date.
This choice is relevant as long term variations of plasma measurements
(usually \Te) are then easily seen.

All JET pulses between $35000<=\JPN<=79853$ are included in the
forthcoming analyses, where \JPN 35000 was produced in May 1995 and \JPN
79853 is the last pulse of C27 in November 2009. Pulse selection
(pulses/timeslices) is done on plasma parameters as detailed
in (\Part{Sec:Obs1_sets}), providing of course that pulses were not
plasma-free and that the ECE and TS diagnostics were operating.

Routines performing the data processing for all diagnostics, including
ECE and TS as well as all other JET data, are constantly improved,
which usually implies some data reprocessing. JET data used and shown in
this study have been taken as stored in JET public database during
Summer 2011 and no specific reprocessing was performed.

\subsection{Magnetic mapping}
\label{Sec:DataDiagnostics_mapping}

Temperature profiles in tokamaks are four dimensional quantities (three
radial, one temporal). The assumption of toroidal symmetry reduces the
radial dimensions to two. Comparing measurements from different
diagnostics, each having a different LOS, requires to map their
geometrical LOS to the midplane plasma radius (or any other
one) by using a magnetic equilibrium model; \efit \cite{Lao_NF_85,
  obrien_NF_92} is used in this paper with a magnetic reconstruction
only constrained by JET magnetic coils.

Discussing the accuracy of JET ECE radial position of all JET data
studied in this paper is definitely out the scope of this paper. The hypothesis of cold
resonance brings a slight overestimation of ECE radial position -- a few
\SI{}{\centi\meter} for usual JET plasma conditions.  Mapping of LOS to the
central midplane affects it as well; more generally, JET ECE radial
position relies on magnetics measurements.

Accuracy of JET magnetic measurements is $1.5\%$ \footnote{Private
  communication from Sergei Gerasimov}. Following Ref\cite{murari-nf-11}
it leads to an accuracy of \SIrange{1}{1.5}{\centi\meter} for the
magnetic geometric boundaries (for $\rho=1$) as computed by
\efit. Magnetics measurements are not yet constraining \efit enough when
reconstructing the magnetic equilibrium's radial position for
central regions of the plasmas, radial accuracy at the plasma core
is not expected\footnote{Private communication from Vladimir Drozdov} to
be better than \hbox{\mytilde5-\SI{10}{\centi\meter}}.

Radial localisation of ECE measurement depends on \field, following
$\Delta r/r = \Delta B/B$ for cold resonance hypothesis, $r$ being the
plasma major radius. This \field-related uncertainty leads to
\hbox{$\delta r$ \mytilde\SI{5}{\centi\meter}} in the plasma
centre. Taking only into account this (quite large) source of radial
uncertainty stresses the complexity of the problem.

It is expected that these large uncertainties affecting the
estimation of JET ECE radial position will not show clear trends of
drifts with time, and that they are more likely to have the same effect
as a low-amplitude white noise on \Te observables
(see (\Part{Sec:DataDiagnostics_meas})). 

Measurement position of \lidar and \hrts are calibrated during JET
shutdown and do not rely on \efit on their LOS; they do though when
mapping to the plasma midplane to be able to compare them to ECE \Te profiles.


\subsection{Comparing \Te measurements}
\label{Sec:DataDiagnostics_meas}

\subsubsection{\Te observables}
\label{Sec:DataDiagnostics_meas_Te}

Different quantities can be defined to reduce the bi-dimensional
profiles to one single value: they are called
\hbox{\emph{observables}} and are designed to quantify the
discrepancies between \mi and \lidar. Some possibilities have been
studied:

\begin{itemize}
\item radial averaging around the plasma centre: $$\maxTem =
  \text{median} \left(\Tem(\rho,t)), |\rho| <0.2\right)$$
\item integrals of $\Tem(\rho)$ over $\rho$ where measurements were
  available for both diagnostics: $$\intTem = \int_{-0.4}^{0.9}\Tem(\rho)
  \text{d}\rho$$
\item \Te values for fixed $\rho$ values: \TeFix{0.5}, \TeFix{0.8} and
  \TeFix{0.9} for, respectively, $\rho=0.5$, 0.8 and 0.9.
\end{itemize}

In order to reduce the uncertainties caused by a non-ideal midplane
mapping (see (\Part{Sec:DataDiagnostics_mapping})), profiles are radially
averaged over the plasma centre, following \maxTe. This method has been
thoroughly tested and it is regarded as robust, satisfyingly dealing with
high- or low-peaked profiles as well as mapping issues around the plasma
centre.

Using \intTe or \TeFix{0.5} instead of \maxTe would lead to very similar
behaviour (variations and amplitudes). Mapping errors and measurement noise
prevent \TeFix{0.8} and \TeFix{0.9} to be give similar results.


\section{Long-term observations of ECE TS ratio}
\label{Sec:Obs1}

\subsection{Definition of ECE TS ratio}
\label{Sec:Obs1_ECE_TS_ratio}

Stability of JET electron temperature is based on observations of \maxTe
as measured by \mi and \lidar, \ie \maxTeECE/\maxTeLIDAR; this ratio is
called \emph{ECE TS ratio}.

The following paragraphs explain why \ece and \hrts are not used to
build an equivalent ECE TS ratio.

\ece data is calibrated for each pulse against \mi, defining
cross-calibration factors $\Temim/\Teecem$: any drift of \mi calibration
could then be checked against these \ece cross-calibration factors,
provided the time stability of \ece calibration is assessed
independently. Despite our efforts, it has not been possible to use
these \ece cross-calibration factors to check \mi stability, as the
former are obviously not stable enough to be used as references. Gaps in
the cross-calibration factors can be explained by the waveguide switches
that allow to select O- or X- microwave mixers that actually have moving
parts; large time drifts observed in these cross-calibration factors
(up to 96 for the latest set-up) do not all show similar features and
each frequency channel appears to be almost independent from the other
ones. 

\ece should therefore be regarded in this paper as an upgrade of \mi,
with higher time and radial resolution. Two points make \mi best suited
than \ece for the ECE TS ratio; first: \mi LOS is closer to the plasma
centre than \ece's; second: \ece cross-calibration might add additional
uncertainty when estimating \Te.

The main reasons to prefer \lidar to \hrts to build the ECE TS ratio
are, first, the late availability of \hrts (Spring 2008) and, then, the
dependence of \hrts calibration to \ece, then \mi, for all the data
presented in this paper.

For those reasons, ECE TS ratio is made of \mi and \lidar data.

\subsection{About data sets}
\label{Sec:Obs1_sets}

Many plasma parameters can affect \Te, the most obvious probably being
the amount of additional power coupled to the plasma. For non purely
Ohmic pulses, a  selection of pulses/timeslices has been made on
$\Ptotm=\PNBIm+\PICRHm$, where \PNBI is the total additional power from
NBI (Neutral Beam Injection) and \PICRH is the total additional power
from ICRH (Ion Cyclotron Radio-frequency Heating). Pulses with LHCD
(Lower Hybrid Current Drive) are not included in the analysis because
this heating system affects ECE signals, especially at the plasma edge.

The following arbitrary separation on $\Ptotm=\PNBIm+\PICRHm$ is then
used:

\begin{itemize}
\item Ohmic pulses with $\Ptotm=$\SI{0}{\mega\watt}
\item ``Low-power'' pulses with
  $\Ptotm$=\SIrange[range-phrase=-,range-units=single]{0}{10}{\mega\watt}
\item ``High-power'' pulses with
  $\Ptotm$=\SIrange[range-phrase=-,range-units=single]{10}{20}{\mega\watt}
\end{itemize}

 Low-power pulses are mostly in L-mode while high-power ones
are mostly in H-mode; no distinction is actually made between both
confinement modes in this paper.

Two data sets are defined:

\begin{itemize}
\item The ``large data set'' spans 15 years of data produced in JET
  (1995-2009) with a broad selection of plasma parameters. This set
  gives an overview of the ECE TS ratio, \ie how \Te
  measurements from ECE (\mi) and TS (\lidar) systems compare with
  time. 


\item The ``reference data set'' spans 13 years of data produced
  in JET (1998-2009) with narrow selections of plasma parameters (\eg
  \field = \SI{2.41}{\tesla}, \Ip = \SI{1.96}{\mega\ampere}); 18
  parameters are checked and selected to be as constant as
  possible. These pulses are then regarded as ``beacon'' or
  ``reference'' pulses, allowing for more precise study of ECE TS
  variations with time.

\end{itemize}



  


\subsection{Large data set: 1995-2009}
\label{Sec:Obs1_wide_95_09}

\subsubsection{Description}
\label{Sec:Obs1_wide_95_09_desc}

JET data has been analysed from 1995 to 2009 for $35000<=\JPN<=79853$
where both \mi ad \lidar were on-line. Data selection is limited on
\field and \Ip with \field limited to \hbox{
  \SIrange[range-units=single]{2}{3}{\tesla}} and \Ip to
\hbox{\SIrange[range-units=single]{1.5}{4}{\mega\ampere}}; and after
additional power separation:


\begin{tabular}{ lrr}
  { Ohmic:}&  15k pulses & 135k timeslices\\
  { Low-power:}&  6k pulses & 23k timeslices\\
  { High-power:}&  3k pulses & 10k timeslices
  \end{tabular}
  
\subsubsection{Observations: 1995-2009}
\label{Sec:Obs1_wide_95_09_obs1}

\begin{figure}
\includegraphics[width=\linewidth]{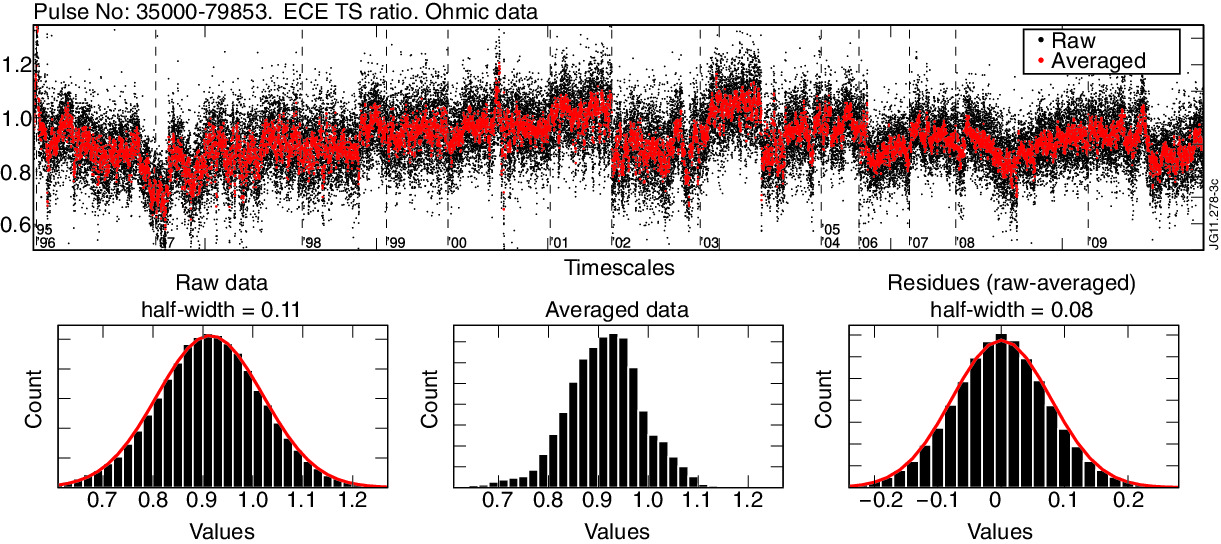}
\caption{\label{Fig2} ECE TS ratio for ohmic pulses, large data set,
  1995-2009.\\ (Top) Data in black, 50-timeslices average in
  red. Years are indicated\\
  (Bottom-left): distribution of ECE TS ratio data, with a half-width of 0.11 and
  Gaussian fit in red.\\
  (Bottom-centre): distribution for 50-timeslices average.\\
  (Bottom, right): distribution of residues, defined as ECE TS ratio
  minus averaged ones, with a half-width of 0.08. Gaussian fit in red}
\end{figure}

In \Fig{2}, the ECE TS ratio is plotted for ohmic conditions. Raw data
ranges from 0.5 to 1.25 and its distribution is Gaussian-like (\Fig{2},
bottom, left) with a mean around 0.95 and a half-width of 0.11. Averaged
data are shown as well, over 50 timeslices that usually represents 5-10
pulses; distribution clearly departs from Gaussian (\Fig{2},
bottom, centre) and residues look like white noise (null mean
Gaussian-like distribution, \Fig{2}, bottom, right) with a 0.08
half-width. Evolution patterns are obviously noticeable on raw and
averaged data.

Focusing on averaged data, ECE TS ratio shows very clear structures
roughly centered on \mytilde 0.95 with \mytilde $20 \%$ temporal
variations (peak-to-peak).

\begin{figure}
\includegraphics[width=\linewidth]{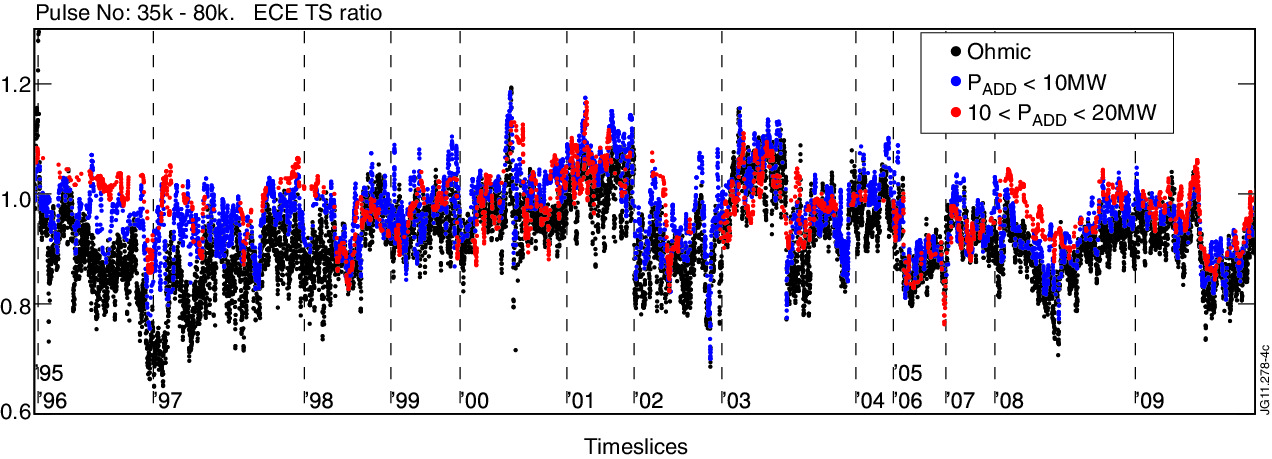}
\caption{\label{Fig3} ECE TS ratio against timeslices for the large data
  set, 1995-2009. Ohmic data in black (15k pulses, 135k timeslices),
  low-power in blue ( \rangePtot{0}{10}, 6k pulses, 23k timeslices),
  high-power in red (\rangePtot{10}{20}, 3k pulses, 10k
  timeslices). Years are indicated by dashed lines.}
\end{figure}

Ohmic, low-power and high-power plasma conditions are plotted in
\Fig{3}, for averaged data as previously defined. The three data sets
mostly exhibit the same trends and a very good agreement after
1998. Compared with the ohmic set, the ECE TS ratio is sometimes
slightly higher for high-power plasma conditions, reaching 10\% between
2007 and 2009 and even 15\% between 1995 to 1998. A slight offset might be
observed between ohmic and non-ohmic data.

Focusing on the ohmic data set, \Fig{2} and \Fig{3} show the same
variations of the ECE TS ratio, that appear to be without clear trends.
Large drops of $\mytilde 20\%$ are observed in 1997, 2001, 2003 and
2009. Numerous increasing or decreasing trends whose amplitudes amount
to 10-20\% can be seen, spanning large numbers of timeslices (thousands
of them, say typically a few months); less sudden variations are also
observed.

Data selection is very broad and spans almost fifteen years. ECE TS
ratio exhibits many sudden changes; explaining the precise causes of
most of those changes is out of the scope of this paper. For instance,
different vessel configurations have been used in JET including the
various divertor, different operation modes,
differently shaped plasmas etc. It has been checked that these changes
of divertor configuration do not correspond with any obvious evolution
of the ratio.


\subsubsection{Observations: C20-C27, 2008-2009}
\label{Sec:Obs1_wide_95_09_obs2}

\begin{figure}
\includegraphics[width=\linewidth]{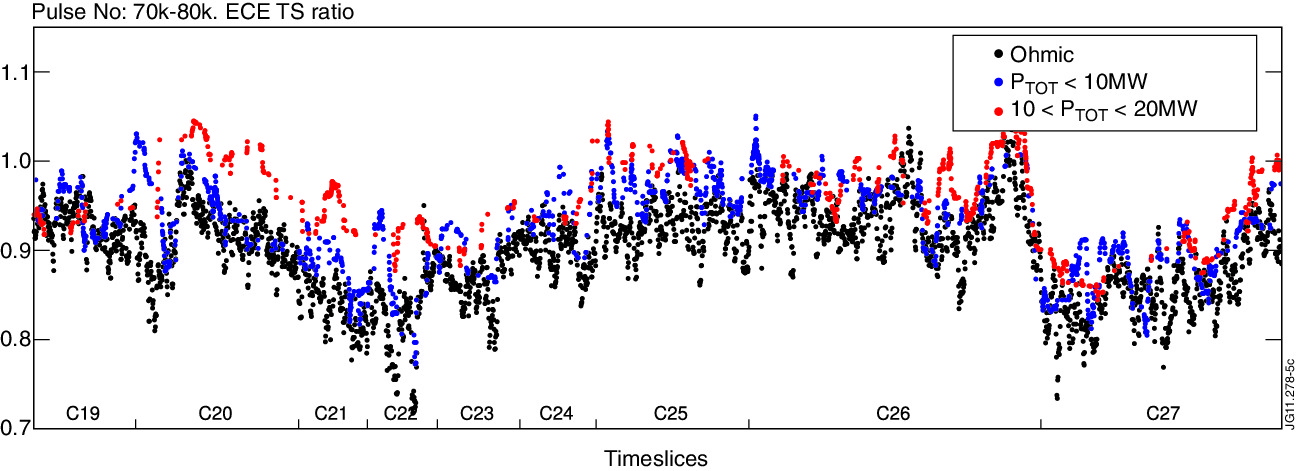}
\caption{\label{Fig4} ECE TS ratio against timeslices for large data set
  limited to C20-C27, 2008-2009, \ie 5800 pulses, 41k timeslices. Ohmic
  data in black low-power in blue ( \rangePtot{0}{10}), high-power in
  red (\rangePtot{10}{20}). Data are averaged over 50 timeslices. JET
  campaigns are indicated.}
 \end{figure}

 ECE TS ratio shown in \Fig{4} is for a time selection limited to the
  JET campaigns between 2008 and 2009 (C20-C27,
 $72000<=\JPN<=79853$). When available, \ece and \hrts data have been
 selected as well -- but not shown in \Fig{4}.

 Like Fig{3}, ECE TS ratio shows \mypm{13}
 variations within [0.8-1.05], with the same strong evidence of
 evolution patterns well outside a \mytilde 8\% near-normal
 scattering. Ohmic, low-power and high-power wide sets show almost
 the same evolutions and levels (ohmic and low-power only) for the
 ratio, whereas the high-power ones have a slightly higher level in some
 case, up to 10\% for C20.


\subsubsection{Key points}
\label{Sec:Obs1_wide_95_09_conc}

This is one of the main conclusion of the paper: between 1995 and 2009,
the (temporal) stability of ECE TS ratio can not be guaranteed, even for
short time-period (say a few months), as shown in \Fig{2} and \Fig{3};
\Te measurements of \mi and \lidar do not compare well, even on
averaged data. This point is regarded as a problematic one from a data
coherence point-of-view, as this ratio is expected to stay around 1.0
with some reasonable margins.

Large-scale drifts of the ECE TS ratio can obviously not been
explained satisfactorily by monotonic drifts of ECE or \lidar
calibrations. This point will be discussed more thoroughly (mostly
in (\Part{Sec:Obs3})).





\subsection{1998-2009 reference pulses}
\label{Sec:Obs1_ref_0809}

ECE TS ratio variations previously detailed for the large data set might
be explained by variations of one or many plasma parameters. In order to
identify possible cause-and-effect links, a first step is to select
pulses with plasma conditions kept as constant as possible.

\subsubsection{Description}
\label{Sec:Obs1_ref_0809_desc}

As such dedicated reference pulses have been neither designed nor
performed through the years, a few sets of plasma
conditions have been chosen and flagged as being used as
\emph{reference pulses}: recovery or cleaning pulses with a long enough  Ohmic
phase (\SI{3}{\second}). Usually performed for the machine to
recover from a disruption, they are usually made on a daily or weekly
basis and do have very similar conditions through the years. These recovery
pulses were not designed to be fully reproducible, \eg the impurities
deposits displaced by the disruption they are recovering from are
expected to affect the plasma composition, but no better pulses sets have
been found. 

Likewise, references pulses were not designed to maintain a
constant \Te.

\begin{table}[b]
  \begin{center}

   \begin{tabular}{ c r r l r}
      \hline
      Signal & &Average & $\pm$ dev. & in \%  \\
      \hline
      \field \hfill [\si{\tesla}]  & &
      2.41   & $\pm$0.01  &(0.4\%)\\
   
      \Ip \hfill [\si{\mega\ampere}] & &
      1.96  & $\pm$\num{1.2e-2} &(0.6\%)\\
      
      \Pohm \hfill [\si{\mega\watt}] & & 
      0.81  & $\pm$0.52 &(65\%)\\
      
      \Wdia  \hfill  [\si{\mega\joule}] &($\star$)&	
      0.90  & $\pm$\num{5.4e-2} & (6\%)\\
      
      $\mathrm{ne_0}$ \hfill [\SI{1e19}{m^{-3}}] & & 
      2.24 & $\pm$0.27 &(12\%)\\

      $\mathrm{P_{e0}}$ \hfill [\si{\kilo\pascal}] & &
      6.80   &$\pm$2.0  &(29\%)\\

      \maxTeECE \hfill [\si{\kilo\electronvolt}] &&
      2.00  & $\pm$0.28 &(14\%)\\
      
      \maxTeLIDAR \hfill [\si{\kilo\electronvolt}] && 
      2.17  &$\pm$0.35  &(16\%)\\
      
      \Zeff \hfill { }  & &
      1.99  &$\pm$0.69   &(35\%)\\

      $\beta_n$  \hfill { }& ($\star$)  & 
      0.38 &$\pm$\num{2.4e-2} & (6\%)\\
      
      \li  \hfill {} & ($\star$) & 
      1.19 &$\pm$\num{3.7e-2} &(3\%)\\

      $\mathrm{q_{95\%}}$  \hfill {} & ($\star$)  &
      3.67   &$\pm$0.21  & (6\%)\\

      plasma volume \hfill [\si{m^3}] &($\star$)& 
      80.7  &$\pm$1.80 &(2\%)\\
      
      triangularity (upper)  \hfill { } & ($\star$)   &
      0.14 &$\pm$\num{8.3e-2}  &(57\%)\\

      triangularity (lower) \hfill { } &($\star$)   & 
      0.18 &$\pm$0.07  &(39\%)\\

      elongation  \hfill { } & ($\star$)  & 
      1.56  &$\pm$0.12  & (7\%)\\

      midplane centre r-pos   \hfill [\si{\meter}] & ($\star$)&	
      2.94 &$\pm$\num{1.1e-2} & (0.3\%)\\

      midplane centre z-pos   \hfill  [\si{\meter}] &  ($\star$)& 
      0.31 &$\pm$\num{2.3e-2}  & (7\%)\\
      \hline
    \end{tabular}
    \caption{ \label{Tab1} Plasma parameters for reference data set,
      limited to $58632<\JPN<79645$, 05/03/03 to 14/10/09, 5702 timeslices, 331
      pulses.\\ ($\star$): data computed by \efit.}
  \end{center}
\end{table}

As detailed in \Tab{1} where typical ranges of main plasma parameters
are listed, 331 reference pulses have been manually selected for
$58300<\JPN<79853$ (2003-2009) corresponding to 5702 timeslices. For
dates prior to 2003, very few pulses could be found with \field =
\SI{2.41}{\tesla} and \Ip = \SI{1.96}{\mega\ampere}.

Some signals are strongly scattered or show non-negligible variations:
Ohmic power (\Pohm), central electron density ($\Nem_0$), central electron
pressure ($\mathrm{P_{e0}}$), effective charge (\Zeff), plasma
triangularities and \maxTe from \mi and \lidar. Besides \maxTe, the only
measured signals showing remarkable trends are the vacuum field
(\mypm{0.4}) with similar variations as the total field, and \Zeff
(\mypm{25}).

\subsubsection{Observations}
\label{Sec:Obs1_ref_0809_obs}

\begin{figure}
  \includegraphics[width=\linewidth]{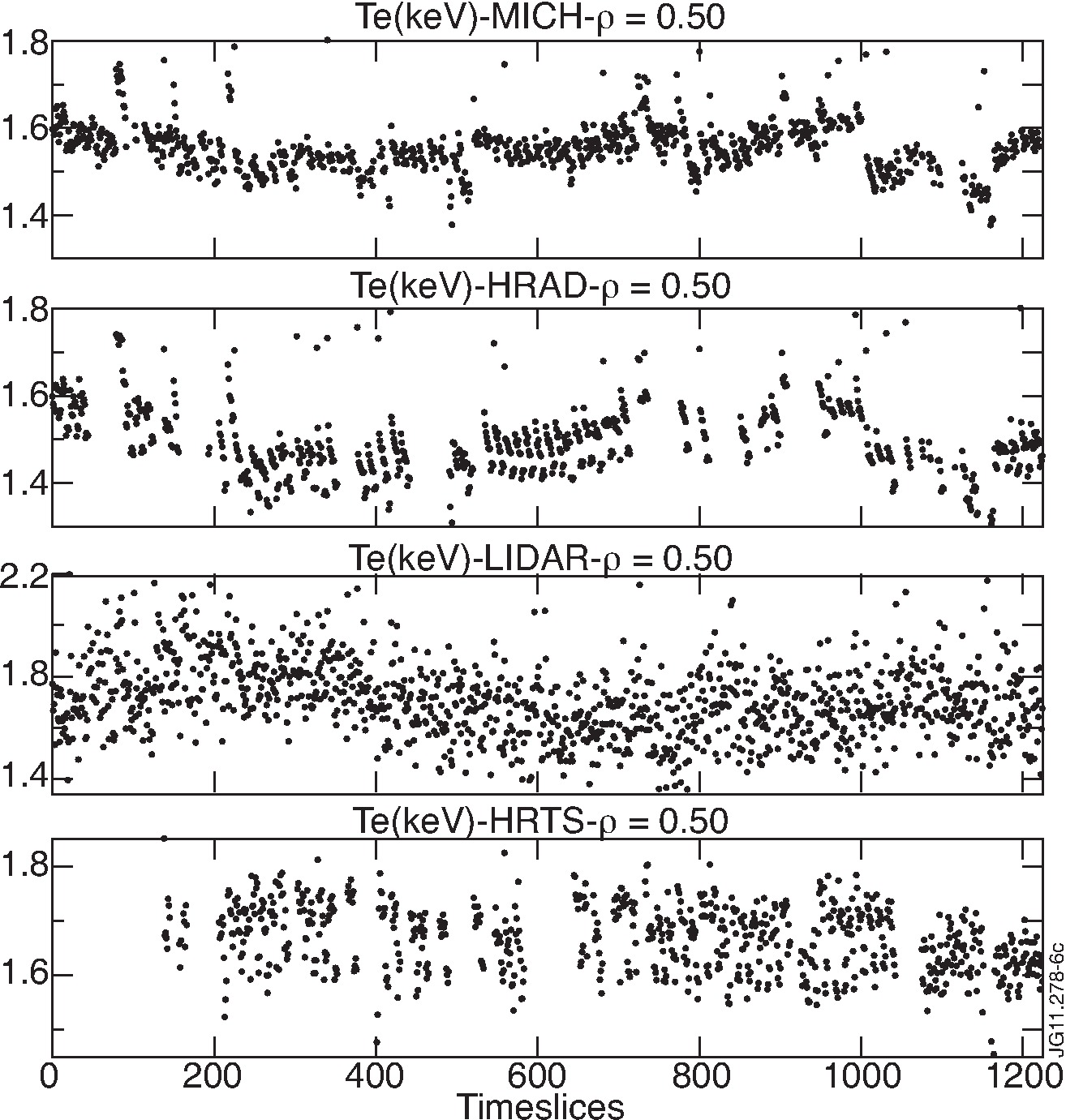}
 \caption{\label{Fig5b} Electron temperature
    measurements  (\si{\kilo\electronvolt}, $\rho=0.5$) against
    timeslices for reference data set as described in \Tab{1}, limited to C20-27
    pulses: 124 pulses and 1227 timeslices selected.  Measurement from different
    diagnostics are shown: \mi, \ece, \lidar and \hrts.  }
\end{figure}

As shown in \Fig{5b}, four diagnostics are available for C20-C27
campaigns: \mi, \lidar, \ece and \hrts, for $\rho=0.5$. Each diagnostic
shows similar variations for different radial positions $\rho= [0.2,
0.3, 0.5]$, even if these variations are not always similar from one
diagnostic to another. Pulse-to-pulse cross-calibration makes \ece
follow \mi very closely, as expected. Even if \hrts is cross-calibrated
against \ece for the whole C20-C27 period (see
(\Part{Sec:DataDiagnostics_diags})), variations of both measurements do
not appear to be linked. Note that \lidar and \hrts exhibit different
variations, even if their measurements are from the same physics basics.


As shown in \Fig{5b}, \Te is obviously not constant for the reference
set: no reference temperature can then be used to assess the stability
of diagnostics. Similarities between the ECE TS ratio and \maxTeECE and
\maxTeLIDAR are to be expected, but depending on the pulse range, trends
have features sometimes linked to \maxTeECE variations and sometimes
\maxTeLIDAR: again, no firm conclusion can be drawn on
the cause of those drifts.


\begin{figure}
\includegraphics[width=\linewidth]{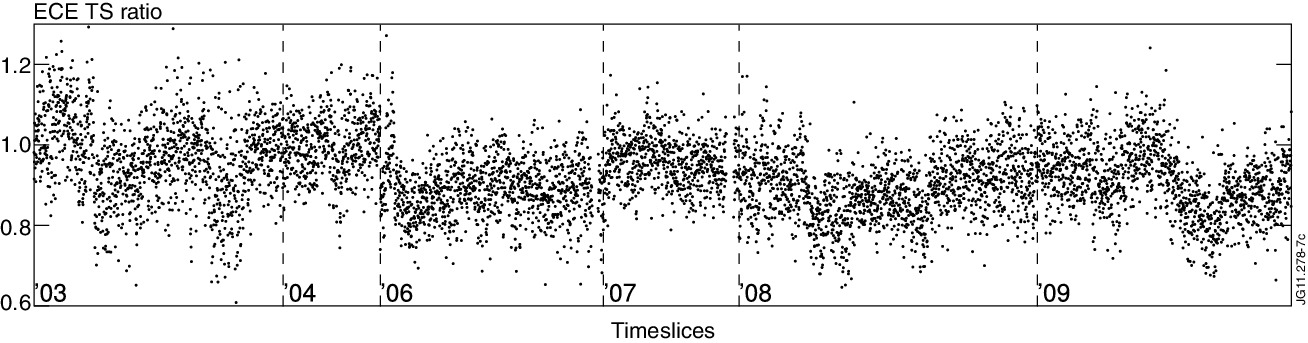}
\caption{\label{Fig5a} ECE TS ratio against timeslices for reference
  data set as described in \Tab{1}. Years are indicated by dashed lines.}
\end{figure}

The ECE TS ratio for reference pulses is plotted in \Fig{5a}. As in
(\Part{Sec:Obs1_wide_95_09_obs1}), residues have a near-normal
(Gaussian-like) distribution similar to the one obtained for the large
data set -- with the mean around 0.95 and non-negligible scattering (=half-width)
of 0.1. Peak-to-peak amplitude is slightly above 30\%.

ECE TS ratio drifts and variations are found to be in agreement with
the large sets (\Fig{3}) and Pearson's cross correlation coefficient
between ratios from both sets is quite high: around 0.7 for $\JPN>54500$
(from 2003 to 2009) as well as for C19-C27 ($\JPN>70000$).

\begin{figure}
\includegraphics[width=\linewidth]{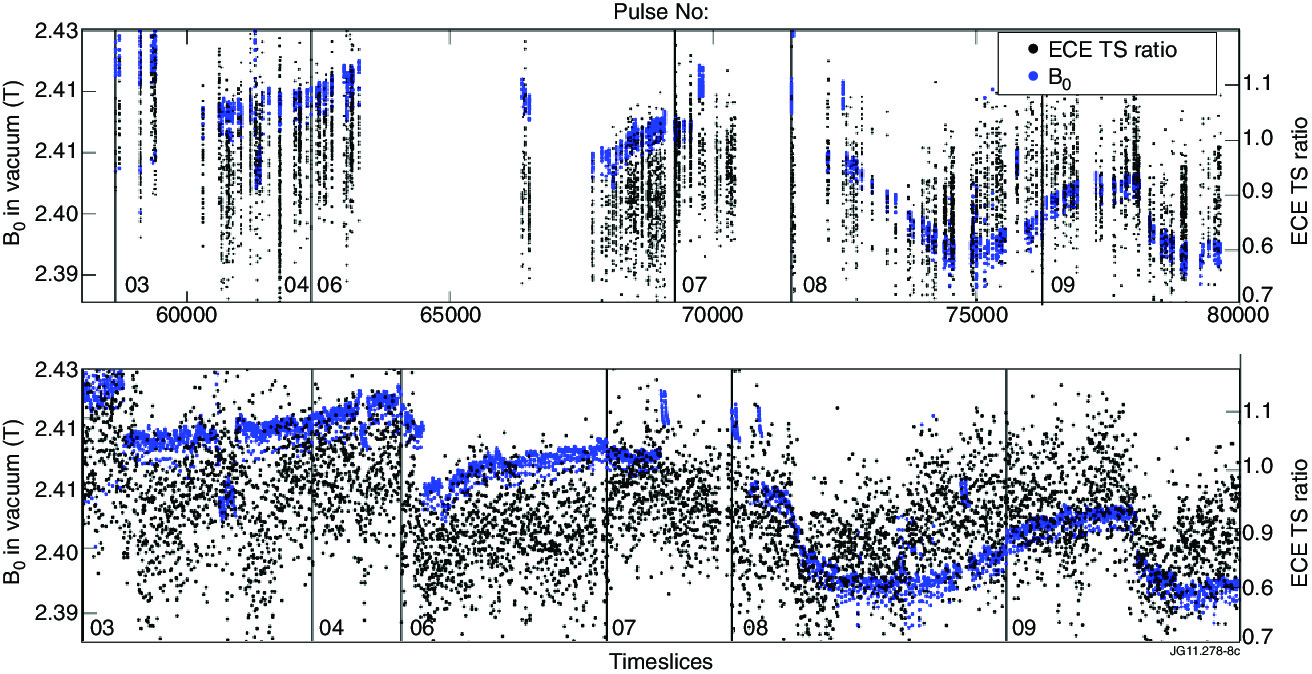}
\caption{\label{Fig6} ECE TS ratio (left y-axis, black) and vacuum
  magnetic field (\field, right y-axis, blue) against \JPN (top) and timeslices
  (bottom) for reference data set as described in \Tab{1}. Years are
  indicated by lines.}
\end{figure}

As stated previously, \field and \Zeff are the only measurements (other
than \Te) showing noticeable trends. \Zeff does not show any variations
similar to the ECE TS ratio, yet. In \Fig{6}, magnetic field (measured
by Rogowskii coils) show large-scale drifts similar to the ones observed
for ECE TS ratio, but the observed \field variations have a very low
amplitude, around \mypm{0.4}; note that these experimental variations
are below the uncertainty level of \kcd integrator measurement (1.5\%, see
(\Part{Sec:DataDiagnostics_mapping})). 

As stated in (\Part{Sec:DataDiagnostics_diags}) and shown in \Fig{1}, \mi
and \lidar LOS are not similar, both missing the plasma center by
\mytilde \SI{5}{\centi\meter} according to \efit reconstruction. For
reference pulses, r- and z-position are observed to move by 
approximately 2-\SI{3}{\centi\meter} at most (see \Tab{1}). Note that from
(\Part{Sec:DataDiagnostics_mapping}) these values are believed to be within
the uncertainties of \efit reconstruction in the plasma center. Even if
those displacements are regarded as real, \Te profiles in usual JET
plasmas as like for the (ohmic) reference pulses are not peaked enough to
explain this 30\% peak-to-peak variations. Furthermore, measurements are
actually volume-averaged for both \mi (usually a few \SI{}{\centi\meter})
and \lidar (\mytilde \SI{1}{\centi\meter}).

\subsubsection{Key points}
\label{Sec:Obs1_ref_0809_conc}

For reference pulses, variations of ECE TS ratio appear to be mostly
disconnected from the variations observed on the 18 plasma parameters
described in \Tab{1}, except for the very low amplitude variations of
\field. 

About the similar long-term drifts shown by \field and ECE TS ratio
(\Fig{6}), variations amplitudes differ from more than one order of
magnitude: \mypm{0.4} for \field and \mypm{15} for ECE TS ratio. Yet
long-term trends of both quantities show a reasonably good
agreement. But \field variations do not affect the computation of
\maxTeECE and \maxTeLIDAR. For ECE diagnostics, frequency-position
mapping adds radial uncertainties following \hbox{$\Delta r/r$ \mytilde
  $\Delta B/B$}: \mypm{0.4} for \field converts then to
$\pm$1-\SI{1.5}{\centi\meter} at the plasma centre
(\hbox{R$\mytilde$\SI{3}{\meter}}), which is below the expected
uncertainties caused by the midplane mapping.

From these two points, the expected impact of \mypm{0.4} \field
variations on \maxTe are considered as negligible -- and
vice versa. ECE TS ratio and \field long-term drifts show then a good
agreement between 2003 and 2009, but no direct cause-and-effect link
between both can be thought of by the authors.

A possible cause for \field variations is described and analysed in the
next section.

\section{Observation of seasonal patterns}
\label{Sec:Obs2}

\subsubsection{Description}

In JET, magnetic field signals (pick-up coils, loops, Rogowskii coils)
used to build \field are all recorded by the same type of 8-channel
isolated integrator (referred as \kcd integrator) used for the principal
magnetic dagnostics.  When those \kcd integrators were installed in
August 1994 there was some concern as to how the inner capacitors would
age and perform with respect to temperature. To estimate the extent of
these effects, a very stable test-signal generator was permanently
installed in \kcd and connected to the 8 channels of a \kcd integrator sample
\cite{Horton_05}. For each pulse, the data from this module are taken
and stored as if they were real coil signals.

\subsubsection{Observations}

The off-plasma signals of \kcd integrator are plotted in
\Fig{7}. Seasonal drifts are observed, as stated in the
report\cite{Horton_05}: \emph{``Up until the JET 2000/2001 shutdown,
  channels 2 to 8 show cyclic variation of about \mypm{0.1}, which
  appear to follow winter-summer pattern''}. The cyclic variations
observed were considered to be temperature effects on these capacitors,
but this point has not been studied further.

\begin{figure}
\includegraphics[width=\linewidth]{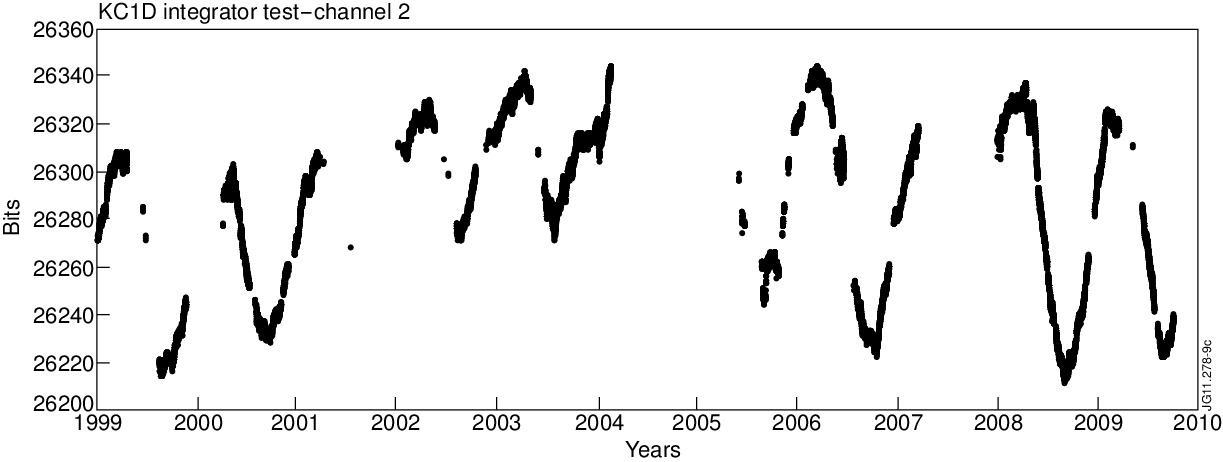}
\caption{\label{Fig7} \kcd integrator test against years. Expected
  values = 26214 bits, drift amplitude = 120 bits = $ \sim 0.2\%$
  variations.}
\end{figure}

As plasma measurements of \field are performed with the same type of
integrators, they are then expected to suffer the same seasonal
variations. As shown in \Fig{8}, \field shows variations quite well
matched by \kcd integrator measurements for the time period mid-2002 to
2009. Amplitudes are quite similar: around \mypm{0.4} for \field and
around \mypm{0.2} for \kcd integrator drifts. The agreement is quite poor
for the 1998-2000 time period.

\begin{figure}
  \includegraphics[width=\linewidth]{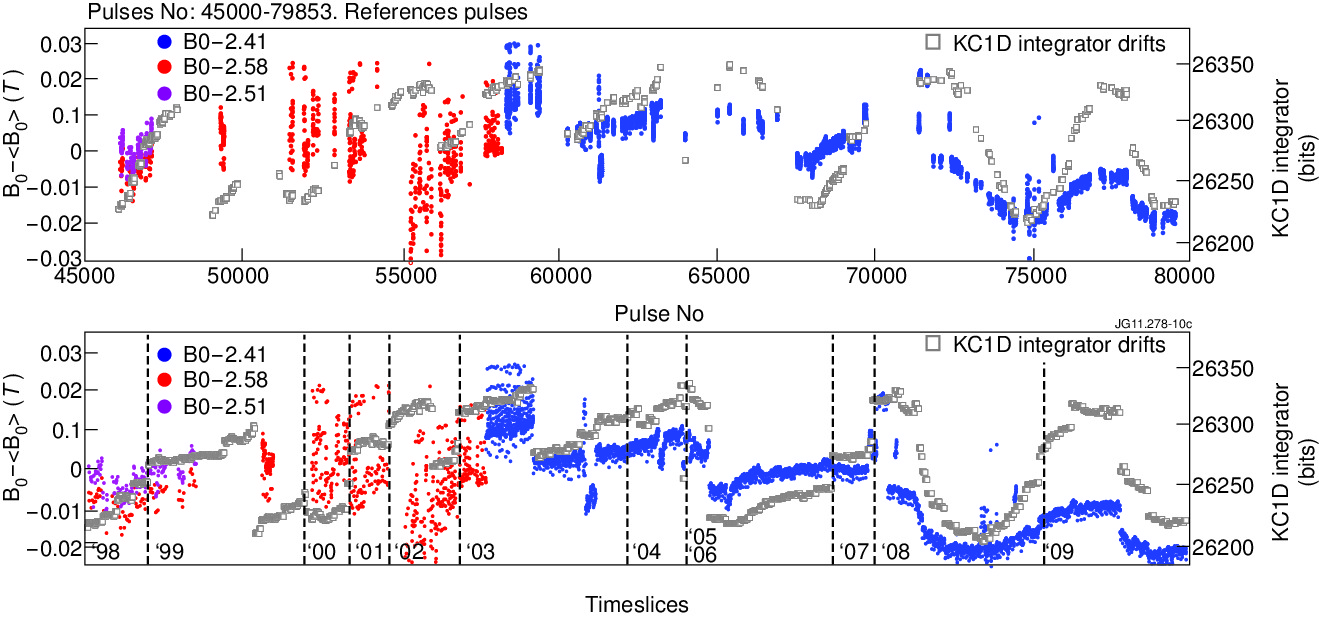}
  \caption{\label{Fig8} Variations of vacuum toroidal field (\field,
    left y-axis, full symbols) and \kcd integrator drifts (right
    y-axis, open symbols, bits-26000) against \JPN (top) and timeslices
    (bottom) for references data set as described in \Tab{1}. Additional
    data sets with similar plasma conditions are shown, for
    \hbox{$45000<\JPN<58632$} with different \field values:
    \SI{2.41}{\tesla} in blue, \SI{2.51}{\tesla} in magenta,
    \SI{2.58}{\tesla} in red. Years are indicated by dashed lines. }
\end{figure}

\subsubsection{Key points}

Given the similarity of \field and \kcd integrator variations, in terms of
amplitude and temporal drifts, it is likely that \field variations
observed in \Fig{6} and \Fig{8} after mid-2002 should not been accounted
for real events taking place in JET plasmas.

Note that those off- and on-plasma variations affecting the field measurements are below
their experimental uncertainty (1.5\%, (\Part{Sec:DataDiagnostics_mapping})).

 \begin{figure}
  \includegraphics[width=\linewidth]{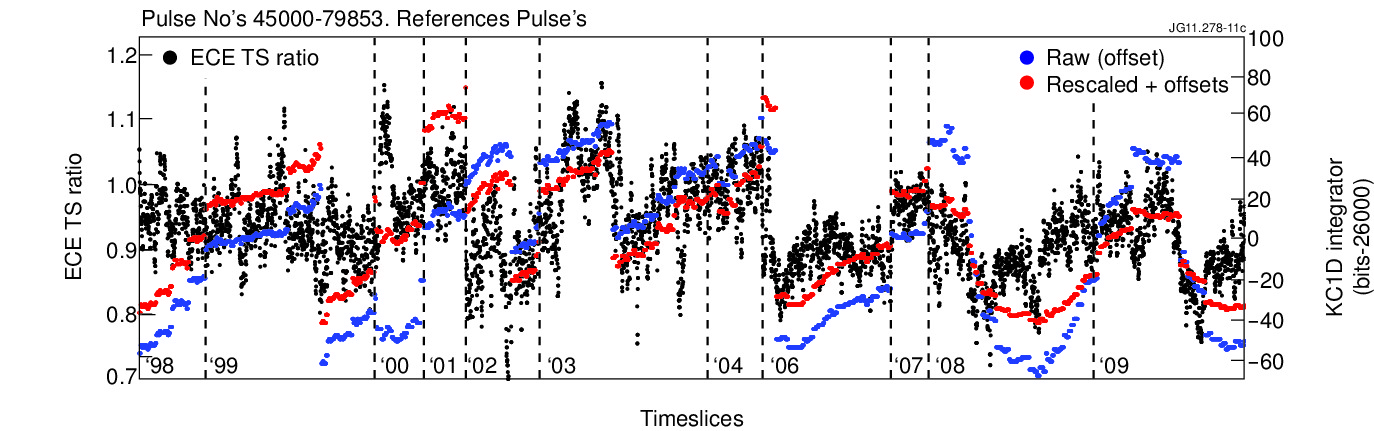}
  \caption{\label{Fig9} ECE TS ratio (left y-axis, black) and \kcd integrator
    drifts (right y-axis, blue) against timeslices for references data
    set as described in \Fig{8}.  \kcd integrator drifts values, slightly
    re-scaled and shifted (offsets $<20$ bits), are in red. Years are
    indicated by dashed lines. }
\end{figure}

As stated in (\Part{Sec:Obs1_ref_0809}) (\Fig{5a}), the only plasma
parameter showing variations correlated with ECE TS ratio for the
reference pulses is \field. The causality of the link between
(on-plasma) ECE TS ratio and (off-plasma) \kcd integrators tests
can not be established from the available data, or the possibility of
a common but still unknown cause having similar effects on both
signals. Seasonal variations of external temperature at JET site are yet
regarded as the main suspect of the long-term drifts of ECE TS ratio
after mid-2002 and since 2009. Little can be said between 1995 and
mid-2002.

Daily variations of JET site temperature might also well
trigger similar effects. The corresponding analysis has been made and
does not show any effect on ECE TS ratio above a $1\%$ uncertainty,
even for the warmer days/months when the night/day temperature
variations reach up to $15\,^{\circ}\!\textrm{C}$ -- quite uncommon for
\hbox{Oxfordshire}\footnote{\url{http://www.metoffice.gov.uk}}. Time
constants considered here are consequently more likely to be of the order
of weeks or months than days.



\section{Long-term stability of Michelson interferometer}
\label{Sec:Obs3}

Another way of explaining the observed variations of ECE TS ratio is to
directly check the variations of \Temi and \Telidar, but even for the
reference data set that is more ``constrained'' than the large data set,
no other \Te measurement can be used as a reference to assess \mi and
\lidar stability. The idea developed in this section is to check \mi
stability with the in-lab (\ie off-plasma) monitoring performed throughout
the years.


\subsection{In-lab measurements}
\label{Sec:Obs3_inlab}

In order to check the stability of \mi, in-lab calibrations are
routinely performed every few months using a calibrated hot source since
the early years of JET (1984); since 1996, the same hot source
(\SI{600}{\degreeCelsius}, built by SPECAC in the 90s) has been
used. Measurements on a cold source (liquid nitrogen) were performed but
not sufficiently frequently (\ie monthly). These measurements were performed by
different \mi operators but as the set-up for in-lab calibrations is
documented and reproducible, it is believed that these measurements were
done in very similar ways.

\begin{figure}
  \includegraphics[width=\linewidth]{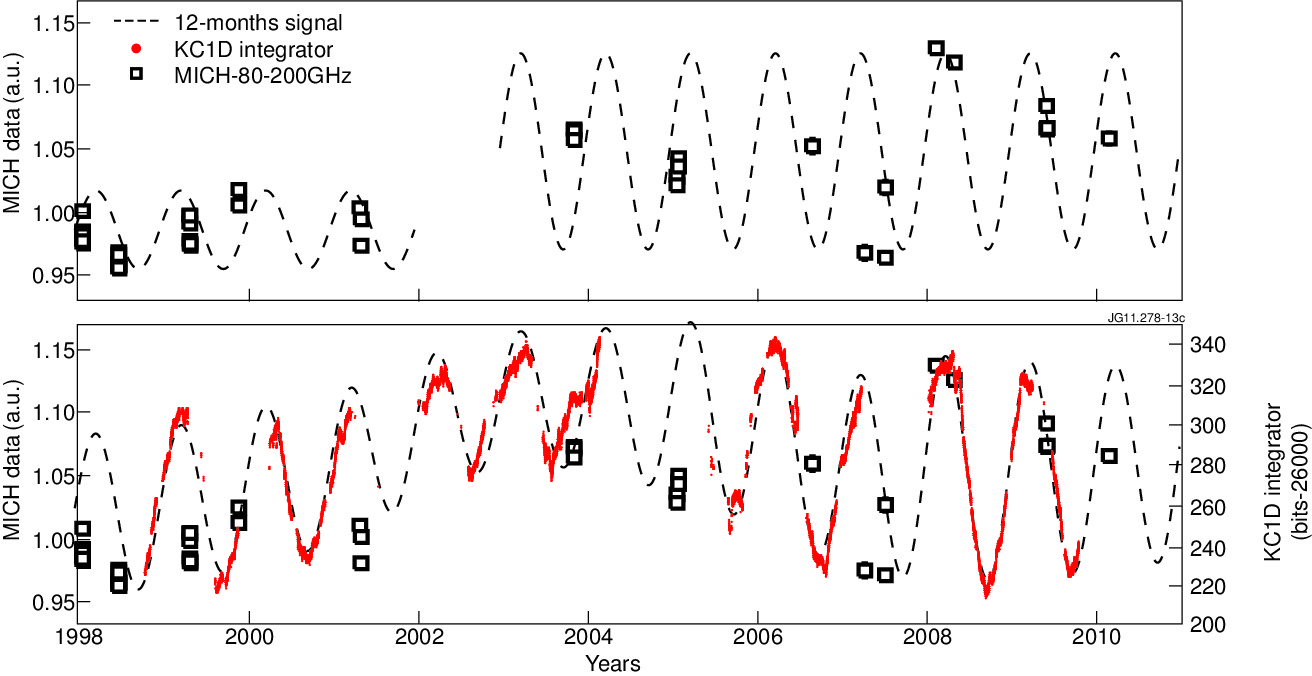}
  \caption{\label{Fig10} \mi in-lab measurements (left y-axis, squares)
    and \kcd integrator drifts (right y-axis, bottom, red points) against time
    for 1998-2010. Synthetic signals with a 12-month period (black
    dashed lines) are added on both plots.}
\end{figure}

Results of \mi in-lab measurements are reported in \Fig{10}. All the
measured interferograms have been reprocessed using the same routines and
the resulting spectrograms have been integrated over the
\SIrange{80}{200}{\giga\hertz} band, which is most frequently used for
measuring \Te with the second ECE harmonic in X-mode.

No cyclic pattern can be easily observed in \Fig{10}. Data scattering is
below \mypm{5} for 1998-2002 and below \mypm{10} for 2003-2010 which is
at least twice lower than the scattering of ECE TS ratio. The scattering
increase of \mi in-lab measurement around 2003 takes place approximately
at the time where ECE TS ratio starts to be well correlated with the
seasonal drifts of the off-plasma magnetics measurements -- as shown in
\Fig{9}. The ECE TS ratio does not show such an increase of scattering.


\subsection{Cryodetector bias voltage}
\label{Sec:Obs3_cryo}

The bias voltage of \mi's cryodetector is provided by batteries
delivering a very stable voltage (\SI{7}{\volt} or \SI{10}{\volt}
depending on the battery model). Sensitivity of the cryodetector to this
bias voltage has been characterised. Taking \SI{10}{\volt} as a
reference, detector sensitivity is slightly increased ($5\%$) for
\SI{7}{\volt} and decreased for lower voltages ($>10\%$ decrease below
\SI{4}{\volt}). Sensitivity variations stay below $5\%$ for a
bias-voltage from \SIrange[range-units=single]{6}{10}{\volt}. This
voltage was monitored on a monthly basis and the consequences of three
serious drops (\SI{<6}{\volt}) have been carefully examined. The
expected impact on \maxTeECE is a (possibly slow) $>10\%$ decrease, then
a steep recovery when batteries are recharged. This signature has not
been observed on JET data; the causality link between such uncontrolled
variations of the \mi cryodetector bias voltage and the drifts of the
ECE TS ratio is not established then.

\subsection{Limits}
\label{Sec:Obs3_limits}

This monitoring includes the \mi system, the cryodetector and some
electronic components. DAQ (hardware) is different for in-lab and plasma
measurements\footnote{A new common DAQ is to be installed for late 2011 or
  early 2012}. The fifty meters of waveguide, the vessel windows and the
antenna are not included in this monitoring.

Components that are included in the monitoring and are the more prone to
suffer from external temperature variations are the cryodetector, which
uses liquid helium, and the electronic amplifiers, which provided a
\SIrange{50}{54}{\decibel} gain for plasma measurements and
\SI{80}{\decibel} for in-lab measurements and in-vessel
calibrations. Their sensitivities to temperature variations ($\mytilde
15\,^{\circ}\!\textrm{C}$) have not been checked yet.

Propagation through the waveguide reduces the signal by
\SIrange[range-units=single]{7}{15}{\decibel}, following an
$\mathrm{F}^2$ scaling between \SI{70}{\giga\hertz} and
\SI{350}{\giga\hertz}. These values are not expected to vary with
temperature. The in-vessel window certainly slightly modifies the EC
signal, but should not cause any seasonal variations. The DAQ's old
electronics might be sensitive to temperature variations. These three
points are possibly causing calibration drifts but their amplitudes are
expected to remain quite small, and they are not expected to introduce
any seasonal variations.

\subsection{Key points}
\label{Sec:Obs3_concl}

This monitoring was historically designed to check the time stability of
\mi between in-vessel calibrations; tracking down such seasonal
variations would require at least monthly measurements, which is not the
case.

The main conclusion is that the available data from \mi monitoring does
not show any 12-months cycles.  Furthermore, data scattering of in-lab
\mi calibration is a factor of two lower than observed on the ECE TS
ratio.


Observed variations of in-lab measurements could come from:

\begin{itemize}

\item Experimental errors in the calibration process, like hot source
  position and temperature and different transmission in the open
  waveguide \ldots but these will not affect \Temi measurement. 

\item \mi calibration drifts that will affect \Temi measurement, for
  example the modification of ageing amplifiers' gain or
  evolution of the components of the cryogenic detector.


\end{itemize}

On the other hand, some other effects described in
(\Part{Sec:Obs3_limits}) and outside the monitoring scope could modify
\Te measurements.

Although available data from \mi lab monitoring does not support the
idea that \mi is mostly responsible for ECE TS ratio variations, lack of
data prevents any firm conclusion to be drawn.

A possibility to definitely track down such calibration drifts would be
to perform a full in-vessel ECE calibration on a monthly basis (at
least), during 1 or 2 years. This is however not feasible on JET.

\section{\Wth estimations}
\label{Sec:Wth}

\newcommand{\Wthmag}{\mathrm{W_{th,mag}}}
\newcommand{\WthECE}{\mathrm{W_{th,\scriptscriptstyle \mi}}}
\newcommand{\WthTS}{\mathrm{W_{th,\scriptscriptstyle \lidar}}}

\subsection{Description}

Plasma thermal energy (\Wth) is estimated from the plasma pressure
profiles, \ie using the kinetic expressions following Ref\cite{garbet-92}:

\begin{alignat*}{1}
  \Wthm &= \Wthme + \Wthmi \\
  \Wthme &= \frac{3}{2} \int{\mathrm{n_e T_e \ dV}}\\
  \Wthmi & \mytilde \frac{7-\Zeffm}{6} \mathrm{ \frac{T_i}{T_e}} \Wthme
\end{alignat*}

Plasma density is taken from \lidar measurements while \Te is measured by
\mi ($\WthECE$) and \lidar ($\WthTS$). Ion temperature is measured by
charge exchange diagnostic\cite{Von_Hellermann_1993}.

Another way to estimate \Wth uses magnetics measurements, defining then
$\Wthmag$:

\begin{center}
  \begin{math}
    \Wthmag = \Wdiam - \frac{3}{2} \Wpetm
  \end{math}
\end{center}

where $\Wpetm$ is the stored energy in the perpendicular component of
the fast particle population, estimated from the computation of power
deposition of ion cyclotron reasonance heating (ICRH)
by \textsc{\footnotesize PION}\cite{eriksson-NF-93,eriksson-PS-95} for
NBI-only plasmas.

Comparing kinetic $\WthECE$ and $\WthTS$ to the diamagnetic $\Wthmag$ provides
information on the measurement accuracy of \Te by \mi and \lidar. \Wth
is then estimated following both ways for purely NBI-heated plasmas from
the wide set, for \PNBI \SI{>5}{\mega\watt}.

\begin{figure}

\includegraphics[width=\linewidth]{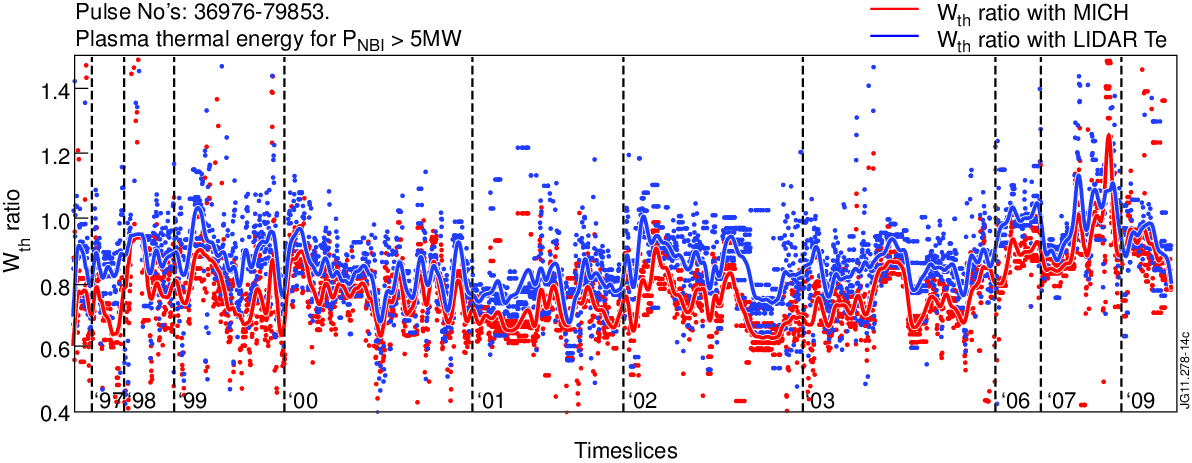}
\caption{\label{Fig_Wth} Ratio of (kinetic) $\Wthm$ to (magnetics)
  $\Wthmag$ with \mi (red) and \lidar (blue) \Te profiles in
  \si{\mega\joule} against timeslices for references data set. Data
  selection from \JPN=36857 to 79853 with \PNBI \SI{>5}{\mega\watt}, 672
  pulses selected, 2604 timeslices. Smaller points represent raw data,
  bigger points represent smoothed data.  Years are indicated by dashed
  lines.}
\end{figure}

\subsection{Observations}

Estimations of $\WthECE/\Wthmag$ and $\WthTS/\Wthmag$ are shown in
\Fig{_Wth}. Large similarities are observed between both kinetics
estimations and they clearly depart from the diamagnetic one, with
variations between -40\% and 10\%.  Most of the time $\WthECE < \WthTS$,
coherent with the observations of ECE TS ratio shown in \Fig{2} and
\Fig{3}: consequently $\WthTS$ estimation is generally slightly closer
to the diamagnetic \Wth than $\WthECE$. No other systematic trend is
observed.

The same exercise has been done for three other cases: first with purely
Ohmic plasma where $\Wthm =\Wdiam$ on a larger pulses database, and then
with a condition on plasma central density for the NBI-only plasmas
(central \hbox{$\Nem_0$ \num{<5e19} \si{m^{-3}}} and \hbox{$\Nem_0$
  \num{>5e19} \si{m^{-3}}}) to check the impact of density on the $T_i$
measurements. These additional cases lead to very similar conclusions as
the ones drawn for the NBI-only case.

\subsection{Key points}

The comparison of kinetic or diamagnetic estimations of
\Wth does not provide any definitive information on the accuracy of \mi or
\lidar measurements. It seems that \mi is slightly under-estimating \Te,
but both of them lead to a \Wth generally 20-30\% below the diamagnetic
\Wth, for both Ohmic and NBI-only plasmas.

\section{ECE TS radial discrepancies}
\label{Sec:Obs4}

\subsection{Description}

Relatively low signal-to-noise ratio and lower radial definition of
\lidar \Te profiles make it impossible to satisfyingly track down
ECE TS position errors. Radially well-defined profiles measured by \hrts
are routinely available since April 2008 (\JPN \mytilde 72000, from
C20). After midplane mapping, they can be compared to the high time
and high (radial) resolution \ece profiles.

A long lasting issue within the JET community is that \ece and \hrts
profiles are not always similar (e.g. as described in
Ref\cite{delaluna-RSI-03, barrera_ppcf_10}); these mismatches are
referred to as \emph{ECE TS radial discrepancies}. As the radial position
of \hrts profiles comes from an independent calibration while ECE
profiles requires a radial-frequency mapping relying on magnetic
reconstruction (as described in (\Part{Sec:DataDiagnostics_mapping})),
radial corrections are usually applied on the ECE side. Profiles
features such as the \Te pedestal in H-mode indicates that at least part of these
mismatches can often be corrected by a radial shift or a slight
modification of the vacuum magnetic field ($<1\%$, less thant the measurement
uncertainty), following $\mytilde k_{corr}$\field and then applying a
non-constant radial shift. Most of these points are carefully addressed
in Ref\cite{delaluna-RSI-03}.

\begin{figure}
\includegraphics[width=\linewidth]{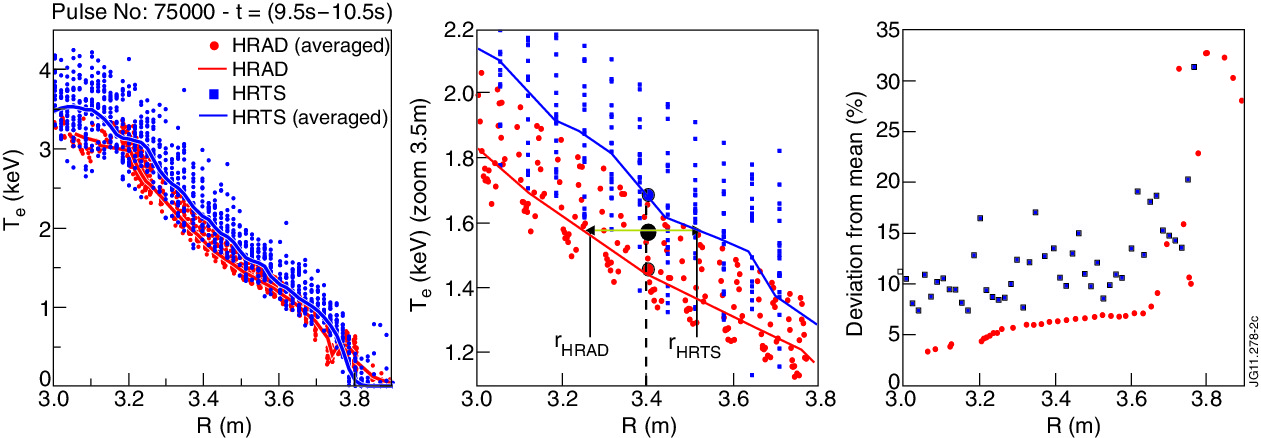}
\caption{\label{Fig1b} ECE TS radial discrepancies for typical JET
  plasma (\JPN 75000, t=9.5-10.5s) for r=\SI{3.5}{\meter}. Plasma
  centre and separatrix are respectively at \mytilde \SI{3}{\meter} and
  \mytilde \SI{3.9}{\meter}. Left: \ece [red] and \hrts
  [blue] \Te profiles; data (points), average
  (lines). Centre: Deviations from time-average for each
  diagnostic. Right: estimation of $\recem - \rhrtsm$ for r=\SI{3.5}{\meter}}
\end{figure}

Observables for ECE TS radial discrepancies are designed in order to
estimate the radial differences between \ece and \hrts
profiles. \Fig{1b} shows how this estimation is done for
r=\SI{3.5}{\meter} for a typical JET pulse. The situation is the same
with $\rho$ profiles. Plasma centre ($\rho\mytilde 0$) is usually missed
by both diagnostics and the edge brings no additional information in
ohmic and L-mode; the focus is then arbitrarily set to the gradient zone
following \hbox{$\median\left(\rhoecem - \rhohrtsm,    0.3<\rho<0.8\right)$}. 

The correction of the ECE TS radial discrepancies, and the various ways
to do so, have been intensively discussed in the JET community for
years. No definite answer has been found yet, because such corrections
must ensure that \ece and \hrts profiles match at the centre, in the
gradient zone and at the plasma edge. Note that the situation worsens
because the degree of the ECE TS mismatch is not always the same for
ohmic, L- and H-mode plasmas.

In this part the possibility of a link between the ECE TS ratio (from
\mi and \lidar) and the ECE TS radial discrepancies (from \hrts and
\ece) is examined.

\subsection{Observations}

\begin{figure}
\includegraphics[width=\linewidth]{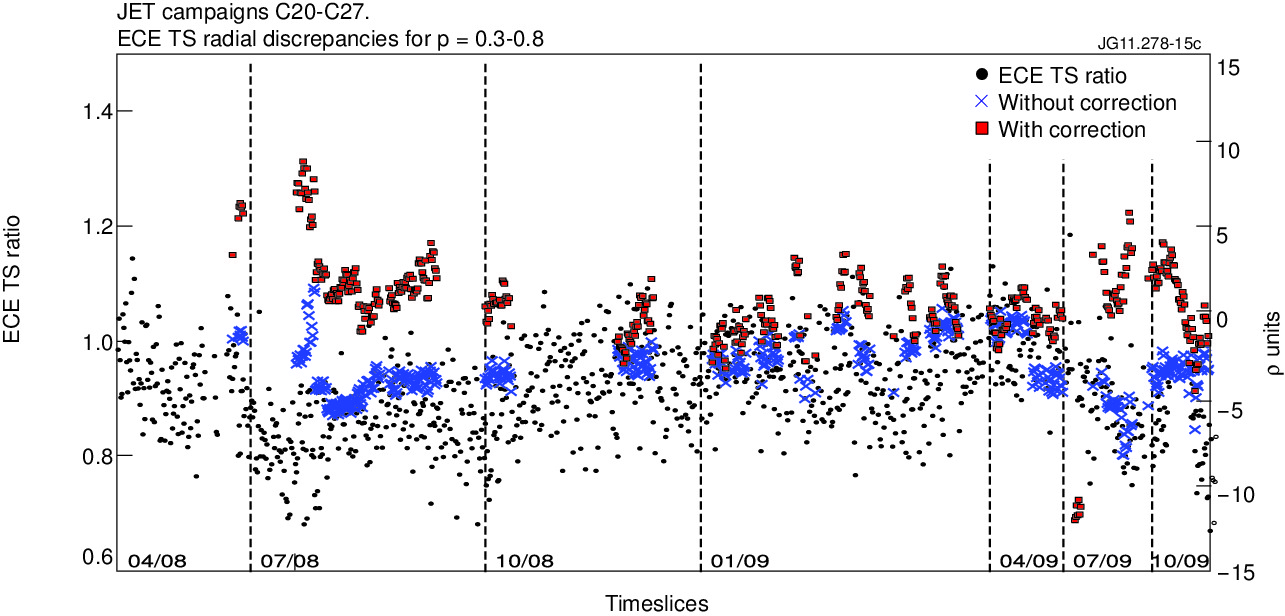}
\caption{\label{Fig13} ECE TS ratio (left y-axis, black points) and ECE
  TS radial discrepancies \hbox{$\median\left(\rhoecem - \rhohrtsm,
      0.3<\rho<0.8 \right)$} (right y-axis, $\rho$ units in $\%$)
  against timeslices for references pulses for C20-C27, where \ece and
  \hrts data are available. Ohmic conditions, 147 pulses, 2878
  timeslices. ECE TS radial discrepancies are non-corrected (cross,
  blue) and corrected (square, red) of the ECE TS ratio variations. Month/years are
  indicated by dashed lines.}
\end{figure}

\begin{figure}
\includegraphics[width=.5\linewidth]{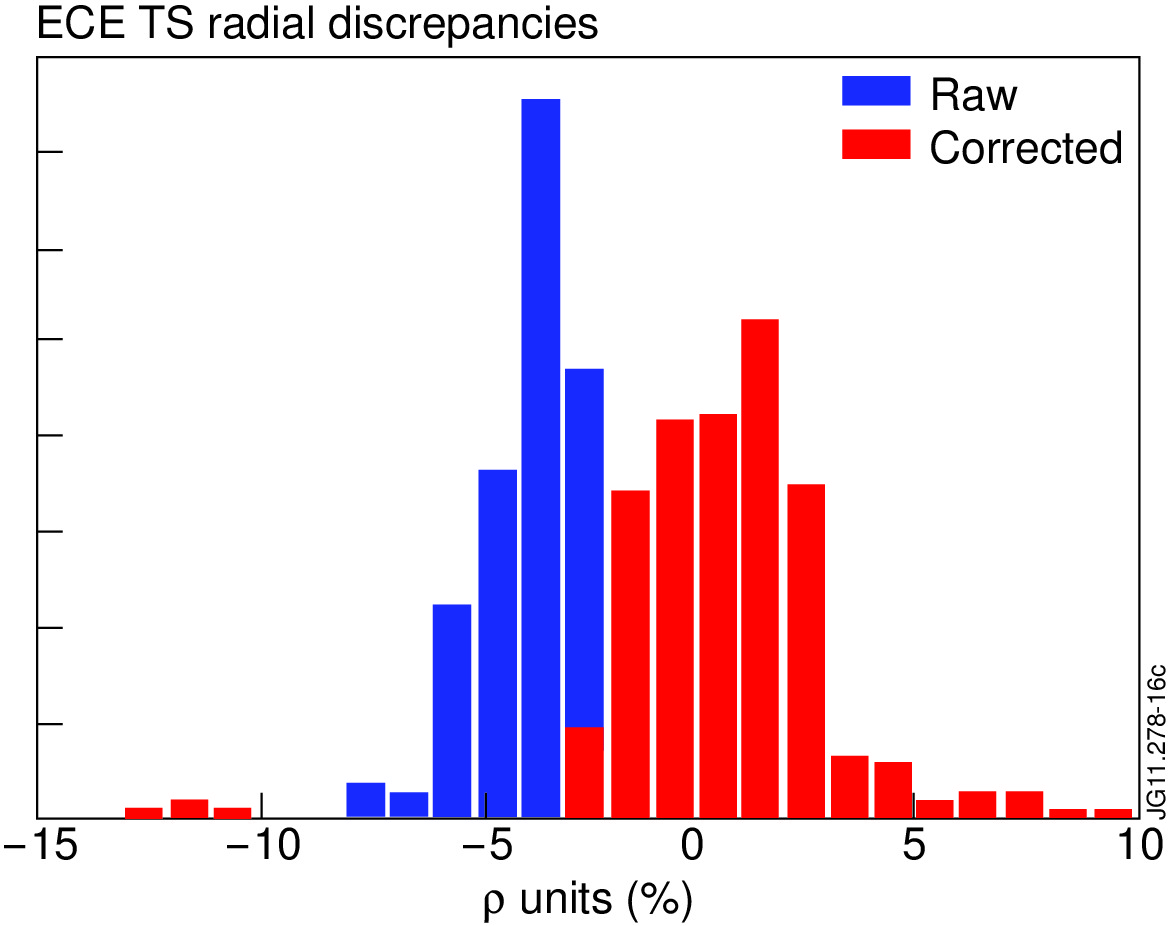}
\caption{\label{Fig14} Distribution of ECE TS radial discrepancies
  ($\rho$ units in $\%$), non-corrected (blue) and corrected (red) of
  the ECE TS ratio variations.  Same pulses/timeslices selection as
  \Fig{13}. }
\end{figure}

In \Fig{13}, profiles mismatches are expressed as amplitude mismatches
(ECE TS ratio, made of \maxTe measurements from \mi and \lidar) and
radial ones (ECE TS radial discrepancies, averaged over $0.3<\rho<0.8$)
for C20-C27 reference pulses where data from all these diagnostics is
available. Only ohmic plasmas are analysed. Both quantities are well
correlated. Radial discrepancies are mostly ranging from $-0.1$ to $0$
($\rho$ units).

Correcting \ece \Te profiles by increasing their values according to the
ECE TS ratio, \ie multiplying \Teece by
\hbox{$\maxTeLIDARm/\maxTeECEm$}, leads to a decrease of the ECE TS
radial discrepancies to a -5\% to 5\% range (\Fig{13}, with \Fig{14} for
the distribution plot). This correction acts like a re-centring of the
radial discrepancies around 0\%; data dispersion is not reduced yet.

\subsection{Key points}

ECE TS radial discrepancies might be explained as an amplitude
shift following the ECE TS ratio -- at least partially. Taking into
account the variations of the ECE TS ratio is regarded as a first-step
correction.

Cross-calibrations of \hrts and \ece against \mi emphasises the role of
\mi, as any possible drifts of \mi calibrations will appear on both ECE
TS ratio and radial discreprancies sides. Possible drifts of \lidar
calibration appear only on the ratio side. Differences between corrected
radial discrepancies and ratio variations are then mostly showing
different behaviours between \lidar central \Te and \hrts \Te in the
gradient zone.

\section{Discussion \& Consequences}
\label{Sec:Dis}

\subsection{Long-term stability of JET \Te measurement}
\label{Sec:Dis_stab}

As defined in (\Part{Sec:DataDiagnostics}), the \maxTe quantity is a
useful means of estimating the central electron temperature of JET
plasmas and of comparing \Te measurements done by diagnostics with
different lines of sight and/or diagnostics that do not always access
the plasma centre.

This paper's main finding is that the observed 15 years of central \Te
measurements (1995-2009) featured in \Fig{3} and \Fig{4} mostly range
between 0.5 and 1.25 (mean around 0.95) and exhibit non negligible
variations through the years, with a 30\% uncertainty observed on
non-averaged ratio (see (\Part{Sec:Obs1_wide_95_09})). Large-scale
structures and patterns are identified as large-scale temporal
variations, amounting up to \mypm{20}. The complexity of the observed
time drifts does not call for a simple explanation. Such variations are
observed as well on more recent JET campaigns (C20-C27) in 2008 and 2009
(\Part{Sec:Obs1_wide_95_09_obs2}).

No measurement for \Te could be used to assess the time stability of any
of the studied diagnostics
(\Part{Sec:Obs1_wide_95_09}, \Part{Sec:Obs1_ref_0809}), even for the
C20-C27 JET campaigns when four diagnostics were available -- only three
being calibrated in a way that their relative time variations are
actually independent: \mi, \lidar and \hrts. Analysis is therefore
limited to relative variations between different measurements.

Following (\Part{Sec:Obs1_ref_0809}), variations of ECE TS ratio are
disconnected from the 18 observed plasma parameters listed in \Tab{1}
for carefuly selected reference pulses, apart from very low variations
(\mypm{0.4}) of the magnetic field \field.

They appear yet to be quite satisfyingly correlated with the
12-month-period patterns shown by off-plasma \kcd integrator signals for
the 2002-2009 time period (\Part{Sec:Obs2}, \Fig{9}). Those variations
of \kcd integrator are believed to be caused by external or ambient
temperature.

 The analysis of the in-lab monitoring of \mi (\Part{Sec:Obs3_inlab}),
 consists of measurements against the same calibrated microwave source
 performed every few months since 1996, in order to check the long-term
 stability of the diagnostic. It is not possible to draw any firm
 conclusion on the link between drifts of ECE TS ratio and in-lab
 measurements performed on \mi, except that this monitoring shows that
 drifts of the ECE TS ratio cannot be explained only by the \mi drifts
 observed in-lab. 

 Regular calibrations of \lidar are regularly performed: measurements of
 spectral functions and detector quantum efficiency are required to
 calculate plasma's \Te. Such calibrations have been done at random
 times of the year, since at least 15 years. These calibrations have
 always been very stable, so the authors would not expect any
 temperature dependance in them.

 Off-plasma monitorings of both \mi and \lidar do not show therefore any
 drifts that have been linked with the observed seasonal drifts of the
 ECE TS ratio.

 For the 1995-2009 time period, JET \Te measured by ECE and TS
 diagnostics show large-scale slow-varying time-dependent variations up
 to \mypm{20}; calibrations drifts are suspected, and quite clear
 correlations are found with seasonal variations of the JET ambient
 temperature since mid-2002 and 2009.

 The authors are not aware of such seasonal drifts between the ECE and
 TS systems installed and operated in any other large sized tokamak. As
 JET systems appear to suffer such sensitivity at an unexpected scale,
 it would be recommended to perform similar systematic checks on the electron
 temperature measurements on other large-sized tokamaks, if
 absolutely calibrated ECE and TS systems are available.

\subsection{Revised uncertainty for overall measurement of JET electron
  temperature}
\label{Sec:Dis_dT}

As long as the cause(s) of the observed variations of ECE TS ratio
remain(s) unknown, it should be considered as uncontrolled uncertainties
affecting JET \Te measurements. The next paragraphs  propose a
revised uncertainty of JET \Te measurements from the experimental
observations presented in this paper, \ie the 1995-2009 time
period. Future measurements are very likely to be similarly affected as
well.  It focuses on central \Te; no better uncertainty is expected for
non-central measurements.

\newcommand{\dT}{\mathrm{\delta T/T}}
\newcommand{\dTcal}{\mathrm{(\delta T/T)_{cal}}}
\newcommand{\dTt}{\mathrm{(\delta T/T)_{t}}}

Experimental uncertainty can be estimated to result from a systematic,
non-time dependent uncertainty $\dTcal$ accounting for absolute
calibration and a time-dependent uncertainty $\dTt$ caused by drifts,
variations or just plain noise, following:

$$
  \dT \sim \sqrt{ {\dTcal}^2 +{\dTt}^2}
$$

Estimated uncertainties of \mi and \lidar calibrations \ie $\dTcal$ are
within 5-10\% (\Part{Sec:DataDiagnostics_diags}). Experimental
large-scale observations reported in this paper represent $\dTt$ and are
estimated at 15-20\%, leading to the following estimation for the
overall uncertainty for JET central \Te measurement:

$$
  \dT \sim 16-22\%
$$

For non-central regions, issues specific to ECE radial
localisation arise, reducing therefore the accuracy of \mi
measurements. \lidar measurements are not affected.

\ece is cross-calibrated against \mi for each pulse, so \Teece
uncertainty can not be lower than \mi data in the plasma center. 
Same remark for non-central \Te.

Absolute calibration of \hrts relies on \ece for C20-C27, therefore, for
this pulse range, it also relies on \mi. As long as no \mi-independent
calibration is used, uncertainties of central and non-central \Tehrts
can not be lower then \mi uncertainties.

Possible drifts of \hrts calibration are not assessed in this paper, but
it can be observed in \Fig{5b} that variations of \hrts measurements for
$\rho=0.5$ might be different to those of \mi or \lidar.  Conclusions for
$\rho=0.2$ and 0.3 are similar. The uncertainty of \hrts data can 
not be estimated to be lower than the estimation for the three other
diagnostics.

It is worth stressing that the confidence intervals resulting from
absolute calibrations of \mi ($\pm{10}\%$) and \lidar ($\pm{5}\%$) --
see (\Part{Sec:DataDiagnostics_diags}) -- are clearly dwarfed by the
time-dependent drifts observed, which are possibly caused by drifts of \mi and
\lidar calibrations.

\subsection{Future work}

Using \mi, \lidar and \hrts as absolutely calibrated diagnostics,
providing independent measurement of the same quantity (\Te), it is
hoped that enough data could be gathered, since C20 (April 2008), in
order to discriminate which diagnostics suffer from calibration
drifts. Few things can be done for the time period before 2008, where
\mi and \lidar were the only calibrated diagnostics.

For the next years of JET operation (2011 and after), the authors
recommend a reference pulse to be designed and performed on a weekly
basis at least for, say, 2 to 3 years. It is of course not possible to
guarantee the perfect stability of the plasma parameters, but
methodically repeating the same pulse will certainly help a lot to track
the stability of the \Te measurements.

In parallel, in-lab monitoring of the diagnostics should be performed
when possible. The frequency of the in-lab monitoring of \mi will be
increased to a weekly or twice per month period. Ways of monitoring \lidar
and \hrts systems should be investigated.

\section{Conclusion}

Between 1995 and 2009, measurement of JET \Te has been carefuly reviewed
using the two main diagnostics available for this time period: \mi
(ECE), \lidar(TS). The overall uncertainty of \Te measurements, based on
these two diagnostics, is estimated to be as high as 16-22\%, mostly
made of large-scale time drifts.

From mid-2002 to 2009, the ratio of temperature measured by the two
diagnostics (namely the ECE TS ratio) shows clear patterns with a
12-months period and a 15-20\% amplitude. These drifts are regarded as
calibration drifts, possibly caused by unexpected sensitivity to unknown
parameters. External temperature on the JET site might is the best
parameter suspected so far. The final cause of such variations has not
been found, nor the way it affects \mi, \lidar and/or (possibly) \hrts
measurements. 

It is not clear whether it is direct impact on one of the diagnostics or
indirect via magnetics measurements, which have been shown to have
seasonal drifts.

Solutions aiming at tracking down these unexpected uncertainties of JET
\Te have been detailed and can be performed during the following JET campaigns
(C28+, after October 2011), including reference pulses designed to be
highly reproducible over time (new ITER-like wall permitting) and specific
off-plasma monitoring of the diagnostics -- when applicable.

Whatever causes these JET \Te drifts, this experimental issue is
regarded as crucial for data quality. Long-term observed ECE TS radial
discrepancies have been shown to be linked to these seasonal
drifts. Tackling this issue and elucidating the cause of these
unexplained seasonal variations is of great importance, impacting to the
quality of JET \Te measurements.

This work was supported by EURATOM and carried out within the framework
of the European Fusion Development Agreement. The views and opinions
expressed herein do not necessarily reflect those of the European
Commission.




\bibliographystyle{aipauth4-1}

%


\end{document}